\pdfoutput=1

\documentclass[aps,prd,%
onecolumn,%
tightenlines,%
eqsecnum,%
showpacs,%
floats,%
nofootinbib,%
longbibliography,
amsfonts,amssymb,amsmath%
]{revtex4-1}

\usepackage{ifthen}
\makeatletter%
\usepackage{xstring}
\newcommand{\pp}{1} 
\IfSubStr{\@classoptionslist}{preprint}%
{\renewcommand{\pp}{1}}%
{\renewcommand{\pp}{0}}%
\makeatother%

\newcommand{\arxiv}{1}

\usepackage{graphicx} 
\usepackage[caption=false]{subfig}             

\graphicspath{{Figures/}}

\usepackage{hyperref}

\newcommand{\ket}[1]{\ensuremath{| {#1} \rangle}}
\newcommand{\bracket}[2]{\ensuremath{\langle {#1} \!\mid\! {#2} \rangle}} 
\newcommand{\ketbra}[2]{\ensuremath{| {#1} \rangle\langle {#2} |}} 
\newcommand{\melt}[3]{\ensuremath{\langle {#1} | {#2} | {#3} \rangle}}

\newcommand{\expct}[1]{\ensuremath{\langle {#1} \rangle}}

\usepackage{dsfont}

\def\Re{\ensuremath{\mathbb{R}}}

\def\IT{\ensuremath{\mathbb{T}}}
\def\Int{\ensuremath{\mathbb{Z}}}

\def\Hphys{{\mathcal H}_{\mathrm{phys}}}
\def\cell{{\mathcal V}}
\def\Vmax{{\mathcal V}}

\def\ow{\mathring{\omega}}
\def\oV{\mathring{V}}

\def\oq{\mathring{q}}

\def\cons{\sqrt{12 \pi G}}
\def\k{\kappa}

\def\rhoc{\rho_{\textit{crit}}}
\def\rhop{\rho_{p}}

\def\lp{l_p}

\def\scri{{\mathcal I}}

\def\v{\nu}
\def\eW{e^{{\scriptscriptstyle\text{WdW}}}}

\def\PsiL{\Psi_{L}}

\def\PsiR{\Psi_{R}}

\def\Psik{\tilde{\Psi}}

\def\PsikL{\tilde{\Psi}_{L}}

\def\PsikR{\tilde{\Psi}_{R}}

\def\PsiW{\Psi^{{\scriptscriptstyle\text{WdW}}}}
\def\PhiW{\Phi^{{\scriptscriptstyle\text{WdW}}}}

\def\PsiWL{\Psi^{{\scriptscriptstyle\text{WdW}}}_{L}}

\def\PsiWR{\Psi^{{\scriptscriptstyle\text{WdW}}}_{R}}

\def\PsiWk{\tilde{\Psi}^{{\scriptscriptstyle\text{WdW}}}}

\def\PsiWkL{\tilde{\Psi}^{{\scriptscriptstyle\text{WdW}}}_{L}}

\def\PsiWkR{\tilde{\Psi}^{{\scriptscriptstyle\text{WdW}}}_{R}}

\def\ThetaW{\Theta^{{\scriptscriptstyle\text{WdW}}}}

\DeclareMathOperator{\sgn}{sgn}

\begin{document}

\preprint{Preprint \#: PI-QG-250}

\title{Dynamical eigenfunctions and critical density\\ in loop quantum cosmology}

\author{David A.~Craig}
\email[]{E-mail: craigda@lemoyne.edu}
\affiliation{%
Perimeter Institute for Theoretical Physics\\
Waterloo, Ontario, N2L 2Y5, Canada\\
and\\
Department of Physics, Le Moyne College\\
Syracuse, New York, 13214, USA}

\date{\today}

\begin{abstract}
We offer a new, physically transparent argument for the existence of the
critical, universal maximum matter density in loop quantum cosmology for the
case of a flat Friedmann-Lema\^{i}tre-Robertson-Walker cosmology with scalar
matter.  The argument is based on the existence of a sharp exponential
ultraviolet cutoff in momentum space on the eigenfunctions of the quantum
cosmological dynamical evolution operator (the gravitational part of the
Hamiltonian constraint), attributable to the fundamental discreteness of
spatial volume in loop quantum cosmology.  The existence of the cutoff is
proved directly from recently found exact solutions for the eigenfunctions for
this model.  As a consequence, the operators corresponding to the momentum of
the scalar field and the spatial volume approximately commute.  The
ultraviolet cutoff then implies that the scalar momentum, though not a bounded
operator, is in effect bounded on subspaces of constant volume, leading to the
upper bound on the expectation value of the matter density.  The maximum
matter density is universal (i.e.\ independent of the quantum state) because
of the linear scaling of the cutoff with volume.  These heuristic arguments
are supplemented by a new proof in the volume representation of the existence
of the maximum matter density.  The techniques employed to demonstrate the
existence of the cutoff also allow us to extract the large-volume limit of the
exact eigenfunctions, confirming earlier numerical and analytical work showing
that the eigenfunctions approach superpositions of the eigenfunctions of the
Wheeler-DeWitt quantization of the same model.  We argue that generic (not
just semiclassical) quantum states approach symmetric superpositions of
expanding and contracting universes.
\end{abstract}

%

\maketitle

\section{Introduction}
\label{sec:intro}

Loop quantized cosmological models generically predict that the ``big bang''
of classical general relativity is replaced by a quantum ``bounce'' in the
deep-Planckian regime, at which the density of matter is bounded by a maximum
density, typically called the ``critical density'' $\rhoc$.  (See Refs.\
\cite{ashsingh11,boj11a} for recent reviews of loop quantum cosmology [LQC]
and what is currently known about these bounds in various models, as well as
references to the earlier literature.)  In most models, the value of this
critical density is inferred from numerical simulations of quasiclassical
states.  So far it has been possible in only a single model -- the exactly
solvable loop quantization, dubbed ``sLQC'' \cite{acs:slqc}, of a flat
Friedmann-Lema\^{i}tre-Robertson-Walker cosmology sourced by a massless,
minimally coupled scalar field -- to demonstrate analytically the existence of
$\rhoc$ for generic quantum states.
In this model, it was shown that $\rhoc \approx 0.41 \rhop$, where $\rhop$ is
the Planck density.%
\footnote{The difference of a factor of 1/2 in the value of $\rhoc$ quoted
here to that in the earliest papers is attributable to the realization in
Ref.\ \cite{awe09a} that the ``area gap'' $\Delta$ of loop quantum gravity
should contain an additional factor of 2 for these models; see footnote 1 of 
Ref.\ \cite{acs:slqc}.  See also footnote \ref{foot:gap}.
} %
As in other models, the bound for the density was first found numerically in
Refs.\ \cite{aps,aps:improved}.  This value was then confirmed and given a
clean analytic proof in Ref.\ \cite{acs:slqc}.

In this paper we offer a new demonstration of
the existence of a critical density in this model with the hope of enriching
the understanding of existing results.  The argument is rooted in a study of
the behavior of the dynamical eigenfunctions of the model's evolution
operator, the gravitational part of the Hamiltonian constraint, based on an
explicit analytical solution for these eigenfunctions found recently in Refs.\
\cite{ach10a,ach10b}.  We will show from this solution that the eigenfunctions
exhibit an exponential cutoff in momentum space that is proportional to the
spatial volume.  This ultraviolet cutoff may be understood as a consequence of
the fundamental discreteness of spatial volume exhibited by these models.  As
a consequence of the cutoff, the quantum operators corresponding to the scalar
momentum and spatial volume approximately commute.  The ultraviolet cutoff
then implies that the scalar momentum -- even though its spectrum is not
bounded -- is in effect bounded on subspaces of constant volume.
The proportionality of the cutoff to the spatial volume then leads to the
existence of a critical density that is universal in the sense that it is
independent of the quantum state.

It has long been understood in the loop quantum cosmology community that the
behavior of the eigenfunctions of the gravitational Hamiltonian constraint
operator is the key to understanding the physics of loop quantum models.  In
particular, the quantum ``repulsion'' generated by quantum geometry at small
volume -- leading to the signature quantum bounce -- was clearly recognized in
the decay of the eigenfunctions at small volume in many examples
\cite{aps,aps:improved}.  (See also e.g.\ Refs.\ \cite{apsv07a,bp08a,kp10a},
among many others.)  It was also recognized numerically that the onset of this
decay as a function of volume depended linearly on the constraint eigenvalue.
(See e.g.\ Refs.\ \cite{apsv07a,bp08a}.)  What is new in this work is the
shift in focus to the behavior of the eigenfunctions as functions of the
continuous variable $k$ labeling the constraint/momentum eigenvalues.  This
allows certain insights that may not be as evident when they are considered as
functions of the discrete volume variable $\v$.  From the exact solutions for
the eigenfunctions of sLQC, we are able to show a genuinely exponential cutoff
in the eigenfunctions as functions of $k$ that sets in at a value of $k$ that
is proportional to the spatial volume, thus confirming and grounding the
numerical observations analytically.  This, of course, is the same cutoff that
manifests as the decay of the eigenfunctions at small volume, considered as
functions of the volume -- this is clearly evident in, for example, Fig.\
\ref{fig:esWedge} -- but seen from a
complementary perspective %
that is in some ways cleaner because of the continuous nature of the 
variable $k$. %
From there, we go on to show how the linear scaling of the cutoff gives rise
to the universal upper limit on the matter density.  To our knowledge, this
connection between the linear scaling of the cutoff on the eigenfunctions and
the existence and value of the universal critical density has not previously
been noted.

Though appealingly intuitive, this argument is essentially heuristic, 
so we supplement it with a new proof in the volume representation of the
existence of a $\rhoc$ in this model.  (The proof of Ref.\ \cite{acs:slqc} is 
in a different representation of the physical operators.)

Thus we are able to offer a clear physical and mathematical account of the
origin and value of the critical density, grounded analytically in the exact
solutions for this model, that complements and confirms extensive numerical
and analytic results extant in the literature.  This perspective may be of
some use in numerical and analytical investigations into the existence of a
critical density in more complex models for which full analytical solutions
are not available.  We expand on this point in the discussion at the end,
after we have developed the necessary details.

As a by-product of the methods employed to reveal the ultraviolet cutoff on
the dynamical eigenfunctions, the semi-classical (large volume) limit of the
eigenfunctions is also obtained from the exact eigenfunctions.  The result
confirms the essence of the result obtained on the basis of analytical and
numerical considerations in Refs.\ \cite{aps,aps:improved}, that the exact
eigenfunctions approach a linear combination of the eigenfunctions for the
Wheeler-DeWitt quantization of the same physical model.  (See also Ref.\
\cite{kp10a}, in which a careful analysis of the asymptotic limit of solutions
to the gravitational constraint arrived at the same result as demonstrated
here from the explicit solutions for the eigenfunctions.)  The domain of
applicability of this approximation is described.  This result is then used to
argue that, in the limit of large spatial volume, generic states in LQC -- not
just quasiclassical ones -- become symmetric superpositions (in a precise
sense to be specified) of expanding and contracting universes.  The symmetry
exhibited in numerical evolutions of semiclassical states -- see e.g.\ Refs.\
\cite{aps,aps:improved,acs:slqc,ashsingh11} -- is therefore not an artifact of
semiclassicality, but a generic property of all states in loop quantum
cosmology.  (Compare Refs.\ \cite{cor-singh08a,kp10a} for analytic results
bounding dispersions of states, showing they remain small on both sides of the
bounce.)


The plan of the paper is as follows.  In Sec.\ \ref{sec:FRWLQC} we summarize
the loop quantization of a flat FLRW spacetime sourced by a massless scalar
field.  Sec.\ \ref{sec:thetaefns} studies the dynamical eigenfunctions
$e^{(s)}_k(\v)$ of the model in detail, exhibiting various explicit forms for
the solutions, and works out the asymptotic behavior of the $e^{(s)}_k(\v)$ in
the limits $|\v|\gg\lambda|k|$ and $\lambda|k|\gg|\v|$, where $\lambda$,
defined in Eq.\ (\ref{eq:lambdadelta}), is related to the LQC ``area gap''.
(The ultraviolet cutoff on the $e^{(s)}_k(\v)$ emerges from this analysis in
Sec.\ \ref{sec:eUV}.)  In Sec.\ \ref{sec:repop} the cutoff is employed to
place bounds on the matrix elements of the physical operators and argue that
the scalar momentum is approximately diagonal in the volume representation.
Section \ref{sec:largeV} applies these results to show that generic states in
sLQC are symmetric superpositions of expanding and contracting Wheeler-DeWitt
universes at large volume.  Finally, Sec.\ \ref{sec:rhocrit} offers an
intuitive argument for the existence of a critical density in this model based
on the UV cutoff for the eigenfunctions, as well as a new analytic proof in
the volume representation.  Section \ref{sec:discuss} closes with some
discussion.

\section{Flat scalar FRW and its loop quantization}
\label{sec:FRWLQC}

In this section we briefly describe the loop quantization of a flat ($k=0$)
Friedmann-Robertson-Walker universe with a massless, minimally coupled scalar
field as a matter source.  The model is worked out in detail in Refs.\
\cite{aps,aps:improved,acs:slqc} (see also Ref.\ \cite{livmb12a}); see Ref.\
\cite{ach10a} for a summary with a useful perspective and Refs.\
\cite{ashsingh11,boj11a} for recent general reviews of results concerning loop
quantizations of cosmological models.

\subsection{Classical homogeneous and isotropic models}
\label{sec:FRW}

The starting point is a flat, fiducial metric $\oq_{ab}$ on a spatial 
manifold $\Sigma$ in terms of which the physical 3-metric is given by
$q_{ab} = a^{2}\,\oq_{ab}$, where $a$ is the scale factor.  The full metric is 
given by
\begin{equation}
g_{ab}  =  -n_a n_b + q_{ab},
\label{4metric}
\end{equation}
where the normal $n_a = -N\, dt_a$ to the fixed (${\mathcal L}_t\oq_{ab}=0$) 
spatial slices is given in terms of a global time $t$ and lapse $N(t)$, so 
that $a=a(t)$.

For the Hamiltonian formulation of the quantum theory spatial integrals over a
finite volume are required.  We may therefore either choose $\Sigma$ to have
topology $\IT^{3}$ with volume $\oV$ with respect to $\oq_{ab}$, or topology
$\Re^{3}$ and choose a fixed fiducial cell $\cell$, also with volume $\oV$
with respect to $\oq_{ab}$.  The choice plays no role in the sequel and we
will proceed in the language of the latter choice.%
\footnote{For some discussion of this point see Sec.\ II.A.1 of Ref.\ 
\cite{ashsingh11}.
} %
The physical volume of $\cell$ is therefore $V = a^{3}\,\oV$.

For a massless, minimally coupled scalar field, after the integration over the
spatial cell $\cell$ has been carried out the classical action is
\begin{equation}
S = \oV \int dt \left\{ 
-\frac{3}{8\pi G} \frac{a\dot{a}^2}{N} + \frac{1}{2}a^3\frac{\dot{\phi}^{2}}{N}
\right\}.
\label{eq:classact}
\end{equation}
The classical Hamiltonian is thus
\begin{equation}
H = \frac{1}{\oV}\left\{-\frac{2\pi G}{3}\frac{N}{a}p_a^2 
    + \frac{1}{2}\frac{N}{a^3} p_{\phi}^2\right\},
\label{eq:classham}
\end{equation}
where $p_a$ and $p_{\phi}$ are the canonical momenta conjugate to the scale 
factor and scalar field.

\begin{figure}[hbt!]
\includegraphics[width=0.75\textwidth]{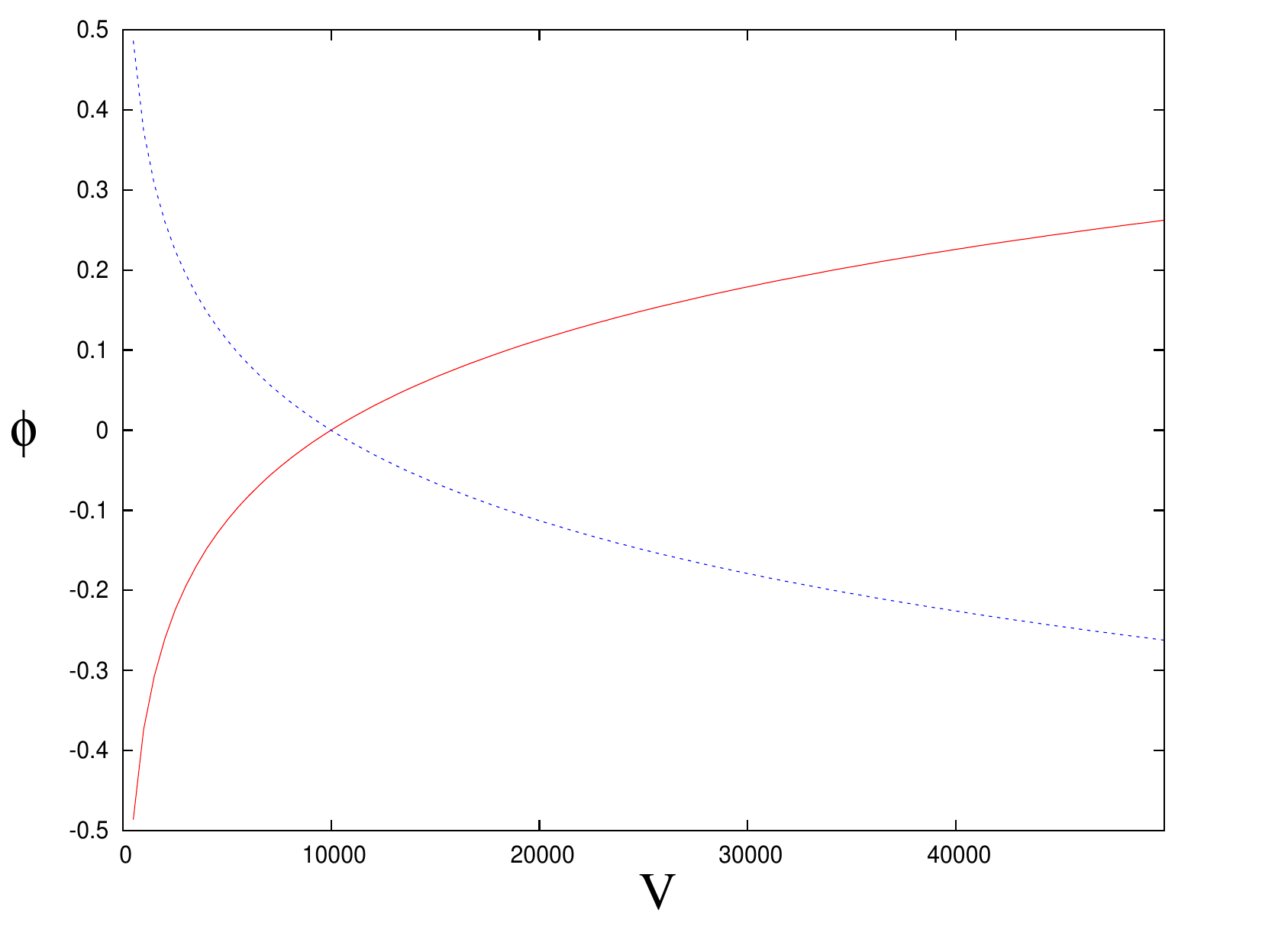}
\caption{Two classical trajectories (Eq.\ (\ref{eq:classtraj})) for a massless
scalar field in a flat homogeneous isotropic universe are shown.
The solid
(red) %
curve corresponds to an expanding branch and the dashed 
(blue) %
curve to the corresponding disjoint contracting branch.  The branches are
singular in the ``past'' and ``future'' given by the internal time $\phi$,
respectively.  (Figure taken from Ref.\ \cite{CS10c}.)
}
\label{fig:classicalsolns}
\end{figure}

Solving Hamilton's equations yields the classical dynamical trajectories, for 
which $p_{\phi}$
is a constant of the motion, and
\begin{equation}
\phi = \pm \, \frac{1}{\cons} \, \ln \left|\frac{V}{V_o}\right| + \phi_o,
\label{eq:classtraj}
\end{equation}
where $V_o$ and $\phi_o$ are constants of integration.  Regarding the value of
the scalar field $\phi$ as an emergent internal physical ``clock'', the
classical trajectories correspond to disjoint expanding ($+$) and contracting
($-$) branches.  The expanding branch has a past singularity (the big bang) in
the limit $\phi \rightarrow - \infty$, and the contracting branch a future
singularity (big crunch) as $\phi \rightarrow +\infty$.  (See Fig.\
\ref{fig:classicalsolns}.)  Note that {\it all} classical solutions of this
model are singular in one of these limits.

Finally, we observe that the matter density $\rho$ on the spatial slices
$\Sigma$ at scalar field value $\phi$ is given in the classical theory by the
ratio of the energy in the scalar field to the volume at that $\phi$:
%
\begin{equation}
\rho|_{\phi} = \frac{p_{\phi}^2}{2 V|_{\phi}^2}.
\label{eq:rhoclass}
\end{equation}
Here $\rho = T_{ab}u^au^b$, where $u^a=(d/d\tau)^a$ and $d\tau=N\,dt$.

\subsection{Loop quantization}
\label{sec:sLQC}

In the quantum theory, following Ref.\ \cite{acs:slqc} we will discuss volume 
in terms of the variable $\v$,
\begin{equation}
\v = \varepsilon \frac{V}{2\pi\gamma l_p^2},
\label{eq:vdef}
\end{equation}
where $\gamma$ is the Barbero-Immirzi parameter, $l_p=\sqrt{G\hbar}$ is the
Planck length (we take $c=1$), and $\varepsilon=\pm 1$ determines the
orientation of the physical triad relative to the fiducial (co-)triad
$\ow^i_a$ determining $\oq_{ab}$ ($=\ow^i_a \ow^j_b\delta_{ij}$) -- see Refs.\
\cite{aps,aps:improved,acs:slqc,ashsingh11}.  Thus $-\infty < \v < +\infty$.
Note that $\v$ is dimensionful.  For comparison to other work, note that $\v =
\lambda\cdot v$, where $v$ is the dimensionless volume variable of Refs.\
\cite{aps,aps:improved},%
\ and
\begin{subequations}
\begin{eqnarray}
\lambda & = & \sqrt{\Delta} \cdot l_p  \label{eq:lambdadelta-a}\\
 & = & \sqrt{4\sqrt{3}\pi\gamma} \cdot l_p.
\label{eq:lambdadelta-b}
\end{eqnarray}
\label{eq:lambdadelta}%
\end{subequations}
Here $\Delta\cdot l_p^2$ is the ``area gap'' of loop quantum gravity.%
\footnote{Note that in earlier work in loop quantum cosmology $\Delta$ was
given as $2\sqrt{3}\pi\gamma$.  However, in Ref.\ \cite{awe09a} it was shown
that $\Delta$ should be taken to have twice that value in homogeneous models.
Since the volume eigenvalues were given in terms of the area gap this
difference does not intrude unduly into those prior results.  The relation
$\v=\lambda\cdot v$ holds so long as one employs the same area gap
consistently throughout.\label{foot:gap}
} %

Remarkably, when the physical model given by Eq.\ (\ref{eq:classact}) is
loop-quantized in these variables, the classical ``harmonic'' gauge choice
$N(t) = a(t)^3$ leads to an \emph{exactly solvable}
quantum theory,%
\footnote{The choice of the harmonic gauge leads to the exact solvability of
the quantum theory because it eliminates inverse factors of $a$ in the
Hamiltonian constraint; cf.\ Eq.\ (\ref{eq:classham}).
} %
referred to as ``sLQC'' (for ``solvable LQC'') \cite{acs:slqc,ashsingh11}.  
One finds that physical states $\Psi(\v,\phi)$ may be chosen to be ``positive 
frequency'' solutions to the quantum constraint,
\begin{equation}
-i\partial_{\phi} \Psi(\v,\phi) = \sqrt{\smash[b]{\Theta_{\v}}}\Psi(\v,\phi),
\label{eq:posfreq}
\end{equation}
where the positive, self-adjoint ``evolution operator'' $\Theta$ (the
quantized gravitational constraint) is given in the $\v$-representation by a
second-order difference operator,%
\footnote{This expression is different from what is found in Refs.\ 
\cite{aps:improved,acs:slqc} because we are using states that carry an 
additional factor of $\sqrt{\lambda/|\v|}$ relative to those states in order 
to simplify the form of the inner product, Eq.\ (\ref{eq:ip}).  Compare, for 
example, Refs.\ \cite{ach10a,ach10b}.
\label{foot:statenorm}
} %
\ifthenelse{\pp=1}{%
\begin{multline}
(\Theta\Psi)(\v,\phi)  =  -\frac{3\pi G}{4\lambda^2} \left\{
\sqrt{|\v(\v+4\lambda)|} |\v+2\lambda| \Psi(\v+4\lambda,\phi) 
- 2 \v^2 \Psi(\v,\phi)  \right. \\
\left.
+ \sqrt{|\v(\v-4\lambda)|} |\v-2\lambda| \Psi(\v-4\lambda,\phi)
\right\}.
\label{eq:theta}
\end{multline}
}{%
\begin{equation}
(\Theta\Psi)(\v,\phi)  =  -\frac{3\pi G}{4\lambda^2} \left\{
\sqrt{|\v(\v+4\lambda)|} |\v+2\lambda| \Psi(\v+4\lambda,\phi) 
- 2 \v^2 \Psi(\v,\phi)
+ \sqrt{|\v(\v-4\lambda)|} |\v-2\lambda| \Psi(\v-4\lambda,\phi)
\right\}.
\label{eq:theta}
\end{equation}
}%
Solutions to the full quantum constraint ($\hat{\mathcal{C}} = 
-[\partial_{\phi}^2 + \Theta]$) therefore decompose into disjoint sectors 
with support on the $\epsilon$-lattices given by $\v=4\lambda n + \epsilon$, 
where $\epsilon\in [0,4\lambda)$ \cite{aps,aps:improved}.  In order not to 
exclude the classical singularity at $\v=0$ from the start, we work 
exclusively on the lattice $\epsilon=0$, so that in this quantum cosmological 
model, the volume is \emph{discrete}:
\begin{equation}
\v = 4\lambda n,   \qquad n\in\Int.
\label{eq:volevals}
\end{equation}
Group averaging yields the physical inner product
\begin{equation}
\bracket{\Psi}{\Phi} = \sum_{\v=4\lambda n} \Psi(\v,\phi_o)^*\Phi(\v,\phi_o)
\label{eq:ip}
\end{equation}
for some fiducial (but irrelevant) $\phi_o$.%
\footnote{See footnote \ref{foot:statenorm}.
} %
According to Eq.\ (\ref{eq:posfreq}), states at different values of the 
scalar field $\phi$ may be mapped onto one another by the unitary evolution
\begin{equation}
\Psi(\v,\phi) = e^{i\sqrt{\smash[b]{\Theta_{\v}}}(\phi-\phi_{o})}\,\Psi(\v,\phi_o),
\label{eq:proppsi}
\end{equation}
It is natural therefore -- though not essential \cite{aps} -- to regard the
scalar field $\phi$ as an emergent physical ``clock'' or ``internal time'' in
which states evolve in this model.  Eq.\ (\ref{eq:proppsi}) shows that the
inner product of Eq.\ (\ref{eq:ip}) is independent of the choice of $\phi_o$,
and is therefore preserved under evolution from one $\phi$-``slice'' to
another.

Finally, we note that in the absence of fermions, the action, dynamics, and
other physics of the model are insensitive to the orientation of the physical
triads \cite{aps,aps:improved,acs:slqc,mbmmo09a}.  We may therefore restrict
attention to the volume-symmetric sector of the theory in which
\begin{equation}
\Psi(\v,\phi)=\Psi(-\v,\phi).
\label{eq:volsym}
\end{equation}

Many further details concerning the quantization of this model and its
observables may be found in Refs.\
\cite{aps,aps:improved,acs:slqc,ashsingh11,livmb12a}.

\subsection{Observables}
\label{sec:observables}

The basic variables in this representation are the scalar field $\phi$ and 
the volume $\v$.  Employing $\phi$ as an internal time, the primary operators 
of interest are the volume, which acts as a multiplication operator,
\begin{subequations}
\begin{eqnarray}
\hat{\v}\,\Psi(\v,\phi) & = & \v\, \Psi(\v,\phi),
\label{eq:volv}\\
\hat{V}\, \Psi(\v,\phi) & = &  2\pi\gamma l_p^2\, |\v|\, \Psi(\v,\phi),
\label{volV}
\end{eqnarray}
\label{eq:volops}%
\end{subequations}
and the scalar momentum $\hat{p}_{\phi}$,
\begin{subequations}
\begin{eqnarray}
\hat{p}_{\phi}\,\Psi(\v,\phi) 
       & = & -i\hbar\,\partial_{\phi}\Psi(\v,\phi)   \label{eq:pphi-a}\\
       & = & \hbar \sqrt{\smash[b]{\Theta_{\v}}}\, \Psi(\v,\phi).  \label{eq:pphi-b}
\end{eqnarray}
\label{eq:pphi}%
\end{subequations}
(In this paper we will not have need of the (exponential of the) momentum
$\boldsymbol{\mathrm{b}}$ conjugate to $\v$ \cite{aps,aps:improved}.)  As in
the classical theory, the scalar momentum $\hat{p}_{\phi}$ is a constant of
the motion -- it obviously commutes with the effective ``dynamics'' given by
$\sqrt{\Theta}$ -- and is therefore a Dirac observable.  The volume $\hat{\v}$
is not, but the corresponding ``relational'' observable $\hat{\v}|_{\phi^*}$
giving the volume at a fixed value $\phi^*$ of the internal time $\phi$ is.
Defining
\begin{equation}
U(\phi) = e^{i\sqrt{\Theta}\phi},
\label{eq:prop}
\end{equation}
the ``Heisenberg'' operator $\hat{\v}|_{\phi^*}(\phi)$ acting on states at 
$\phi$ is given by
\begin{equation}
\hat{\v}|_{\phi^*}(\phi) = U(\phi^*-\phi)^{\dagger}\hat{\v}U(\phi^*-\phi),
\label{eq:volrelnl}
\end{equation}
so that, for example, the physical volume 
$\hat{V} =  2\pi\gamma l_p^2\, |\hat{\v}|$
of the cell $\cell$ at $\phi^*$ is given by the operator
\begin{equation}
\hat{V}|_{\phi^*}(\phi)\, \Psi(\v,\phi) = 
2\pi\gamma l_p^2 e^{i\sqrt{\smash[b]{\Theta_{\v}}}(\phi-\phi^*)}|\v|  \Psi(\v,\phi^*).
\label{eq:volrelnlphys}
\end{equation}
It is straightforward to verify that $\hat{p}_{\phi}$  
and $\hat{V}|_{\phi^*}(\phi)$ commute with $U(\phi)$, and are therefore Dirac
observables.

\section{Eigenfunctions of the evolution operator}
\label{sec:thetaefns}

General physical states $\Psi(\v,\phi)$ may be readily expressed in terms of 
the eigenfunctions of the dynamical evolution operator $\Theta$ -- the 
gravitational part of the Hamiltonian constraint -- given by
\begin{equation}
\Theta_{\v} e_k(\v) = \omega_k^2\, e_k(\v),
\label{eq:thetaefn}
\end{equation}
where
\begin{subequations}
\begin{eqnarray}
\omega_k & = & \sqrt{12\pi G}\, |k|
\label{eq:omegak-a}\\
 & \equiv & \k\, |k|,
\label{eq:omegak-b}
\end{eqnarray}
\label{eq:omegak}%
\end{subequations}
and $-\infty < k < \infty$ is a dimensionless number labelling the 2-fold 
degenerate eigenvalues.  Restricting to the symmetric lattice $\v=4\lambda n$ 
and physical states which satisfy Eq.\ (\ref{eq:volsym}), we usually choose 
to work with a symmetric basis of eigenfunctions $e^{(s)}_k(\v)$ which 
satisfy $e^{(s)}_k(\v) = e^{(s)}_k(-\v)$.  In terms of these physical states 
may be expressed simply as
\begin{equation}
\Psi(\v,\phi) = \int_{-\infty}^{+\infty}dk\, 
   \tilde{\Psi}(k)\, e^{(s)}_k(\v)\, e^{i\omega_k\phi}.
\label{eq:Psiexpn}
\end{equation}
For normalized states, $\sum_{\v=4\lambda n}|\Psi(\v,\phi)|^2 = 1$,
\begin{equation}
\int_{-\infty}^{+\infty}dk\, |\tilde{\Psi}(k)|^2 = 1.
\label{eq:Psinormk}
\end{equation}

Explicit analytic expressions for the eigenfunctions of $\Theta$ have recently
been found \cite{ach10a,ach10b}.  The symmetric eigenfunctions will eventually
be expressed in terms of the primitive eigenfunctions \cite{ach10a}
\begin{subequations}
\begin{eqnarray}
e_{0}(\v) & = & \delta_{0,\v}
\label{eq:e0-a}\\
e_k(\v) & = & A(k) \sqrt{\frac{\lambda|\v|}{\pi}}
     \int_{0}^{\pi/\lambda} db\, e^{-i\frac{\v b}{2}} 
     e^{ik\ln(\tan\frac{\lambda b}{2})}
       \qquad (k\neq 0),
\label{eq:ek-b}
\end{eqnarray}
\label{eq:ek}%
\end{subequations}
where $A(k)$ is a normalization factor which for consistency will always be 
chosen to be  
\begin{equation}
A(k) = \frac{1}{\sqrt{4\pi k\sinh(\pi k)}}.
\label{eq:Ak}
\end{equation}
The functions $e_k(\v)$ have support on both positive and negative $\v$ and 
are not symmetric in $\v$.  It is convenient to seek linear combinations 
$e^{\pm}_{k}(\v)$ of $e_{k}(\v)$ and $e_{-k}(\v)$ which have support only for 
$\v \gtrless 0$.   The correct combinations turn out to be \cite{ach10a}
\begin{equation}
e^{\pm}_{k}(\v) = 
  \frac{1}{2}\left\{e^{\pm\frac{\pi k}{2}} e_{k}(\v) + e^{\mp\frac{\pi k}{2}} e_{-k}(\v)\right\}. 
\label{eq:epmkdef}
\end{equation}
Clearly $\sum_{\v}e^{\pm}_{k}(\v)^*e^{\mp}_{k'}(\v)=0$.  
The choice of $A(k)$ in Eq.\ (\ref{eq:Ak}) corresponds to the normalization 
\begin{equation}
\sum_{\v=4\lambda n} e^{\pm}_{k}(\v)^{*} e^{\pm}_{k'}(\v) = \delta^{(s)}(k,k'),
\label{eq:epmkcompletek}
\end{equation}
where $\delta^{(s)}(k,k')$ is the symmetric delta distribution 
\begin{equation}
\delta^{(s)}(k,k') = \frac{1}{2}\left\{ \delta(k,k') + \delta(k,-k')\right\}.
\label{eq:deltasymm}
\end{equation}
(In contrast to Ref.\ \cite{ach10a}, we choose to work with the full range of 
$k$, $-\infty < k < \infty$.  This leads to the second delta function 
appearing in Eq.\ (\ref{eq:epmkcompletek}) relative to Eq.\ (C13) of that 
reference.  Since as we will see these functions are symmetric in $k$, the 
two approaches are of course equivalent, but do lead to some differences in 
choices of normalization.)

The $e^{\pm}_{k}(\v)$ can be given explicitly as \cite{ach10a}
\begin{subequations}
\begin{eqnarray}
e^{\pm}_{k}(\v) & = & A(k) \sqrt{\frac{\pi|\v|}{\lambda}} 
  I(k,\pm\v/4\lambda)
\label{eq:epmkexplicit-a}\\
 & = & A(k) \sqrt{\frac{\pi|\v|}{\lambda}} I(k,\pm n),
\label{eq:epmkexplicit-b}
\end{eqnarray}
\label{eq:epmkexplicit}%
\end{subequations}
recalling $\v=4\lambda n$.  Here $I(k,n)=0$ for $n<0$, and for $n\geq 0$ is 
given by%
\footnote{Compare Ref.\ \cite{ach10a}, Eq.\ (C8) and Ref.\ \cite{ach10b}, Eq.\ 
(B4).
} %
\begin{subequations}
\begin{eqnarray}
I(k,n) & = & ik \sum_{m=0}^{2n} \frac{1}{m!(2n-m)!}\prod_{l=1}^{2n-1}(ik+m-l)
\label{eq:Ikn-a}\\
 & = & -ik \frac{\Gamma(2n-ik)}{\Gamma(1+2n)\Gamma(1-ik)} 
    \,_2F_1(ik,-2n;1-2n+ik;-1),
\label{eq:Ikn-b}
\end{eqnarray}
\label{eq:Ikn}%
\end{subequations}
where the second form follows from the first by simple manipulations of the 
definition of the hypergeometric function $_2F_1(a,b;c;z)$ \cite{AS64}.  

We will discuss the properties of $I(k,n)$ in detail later.  For now, note 
from Eq.\ (\ref{eq:epmkexplicit}) that
\begin{equation}
e^{\pm}_{k}(-\v) = e^{\mp}_{k}(\v),
\label{eq:epmksv}
\end{equation}
and from Eq.\ (\ref{eq:epmkdef}) that 
\begin{equation}
e^{\pm}_{-k}(\v) = e^{\pm}_{k}(\v).
\label{eq:epmksk}
\end{equation}
Given the symmetry relation Eq.\ (\ref{eq:epmksv}) it is clear that the 
symmetric eigenfunctions $e^{(s)}_k(\v)$ are finally%
\footnote{Notice that restricting to the volume-symmetric sector of the 
theory, Eq.\ (\ref{eq:volsym}), on the $\epsilon=0$ lattice lifts the 2-fold 
degeneracy of the eigenvalues $\omega_k$ to a single eigenvector for each $k$ 
\cite{aps:improved}.
} %
\begin{subequations}
\begin{eqnarray}
e^{(s)}_k(\v) & = & \frac{1}{\sqrt{2}}\left\{ e^+_k(\v) + e^-_k(\v)\right\}
\label{eq:eskdef-a}\\
 & = & A(k) \sqrt{\frac{\pi|\v|}{2\lambda}}\ I(k,|\v|/4\lambda)
\label{eq:eskdef-b}\\
 & = & \sqrt{\frac{|n|}{2|k\sinh(\pi k)|}}\ I(k,|n|).
\label{eq:eskdef-c}
\end{eqnarray}
\label{eq:eskdef}%
\end{subequations}

The following symmetry properties may be verified:%
\footnote{Eq.\ (\ref{eq:esksv}) follows from Eq.\ (\ref{eq:eskdef}).  Eq.\ 
(\ref{eq:esksk}) follows from Eq.\ (\ref{eq:epmksk}).  Eq.\ 
(\ref{eq:eskRe}) follows from a study of $I(k,|n|)$ or the argument to follow.
} %
\begin{subequations}
\begin{eqnarray}
e^{(s)}_{k}(-\v) & = & e^{(s)}_{k}(\v)
\label{eq:esksv}\\
e^{(s)}_{-k}(\v) & = & e^{(s)}_{k}(\v)
\label{eq:esksk}\\
e^{(s)}_{k}(\v)^* & = & e^{(s)}_{k}(\v).
\label{eq:eskRe}
\end{eqnarray}
\label{eq:esks}%
\end{subequations}
The $e^{(s)}_k(\v)$ with $A(k)$ chosen as in Eq.\ (\ref{eq:Ak}) then satisfy 
the completeness relations
\begin{subequations}
\begin{eqnarray}
\sum_{\v=4\lambda n} e^{(s)}_{k}(\v)^* e^{(s)}_{k'}(\v) & = & \delta^{(s)}(k,k')
\label{eq:eskcompletek}\\
\int_{-\infty}^{+\infty}dk\, e^{(s)}_{k}(\v) e^{(s)}_{k}(\v')^* & = & 
\delta^{(s)}_{\v,\v'},
\label{eq:eskcompletev}
\end{eqnarray}
\label{eq:eskcomplete}%
\end{subequations}
where the symmetric Kronecker delta is defined analogously to Eq.\ 
(\ref{eq:deltasymm}).  (The domains of these expressions are understood to be 
even functions of $k$ and $\v$, respectively.)  The symmetrized deltas arise 
because the functions $e^{(s)}_k(\v)$ are symmetric in both $\v$ and $k$.%

An expression for $e^{(s)}_k(\v)$ we will find useful later is
\begin{equation}
e^{(s)}_k(\v) = \frac{\cosh(\pi k/2)}{\sqrt{2}}
   \left\{ e_{k}(\v) + e_{-k}(\v)  \right\},
\label{eq:eskek}
\end{equation}
which follows from Eqs.\ (\ref{eq:eskdef-a}) and (\ref{eq:epmkdef}).   As a 
useful aside, note it is easy to see from Eq.\ (\ref{eq:ek}) that 
$e^{(s)}_{k}(\v)^*  =  e^{(s)}_{-k}(-\v)$.  Additionally, the change of 
variable $b' = -b+\pi/\lambda$ in Eq.\ (\ref{eq:ek}) -- remembering 
$\v=4\lambda n$ -- reveals that
\begin{equation}
e_{k}(-\v) = e_{-k}(\v).
\label{eq:ekvsvk}
\end{equation}
Thus $e_k(\v)^* = e_{-k}(-\v) = e_k(\v)$, and both $e_k(\v)$ and 
$e^{(s)}_k(\v)$ are therefore real.

This completes the catalog of properties of the eigenfunctions we will
require.  We now describe the behavior of the functions $e^{(s)}_k(\v)$ we
seek to explain in the sequel.  
The results of the analysis will confirm and complement the understanding of
earlier numerical and analytical work arrived at prior to the discovery of the
exact solutions for this model.

\begin{figure}[hbtp!]
\subfloat[$n=20$]{
\includegraphics[width=0.50\textwidth]{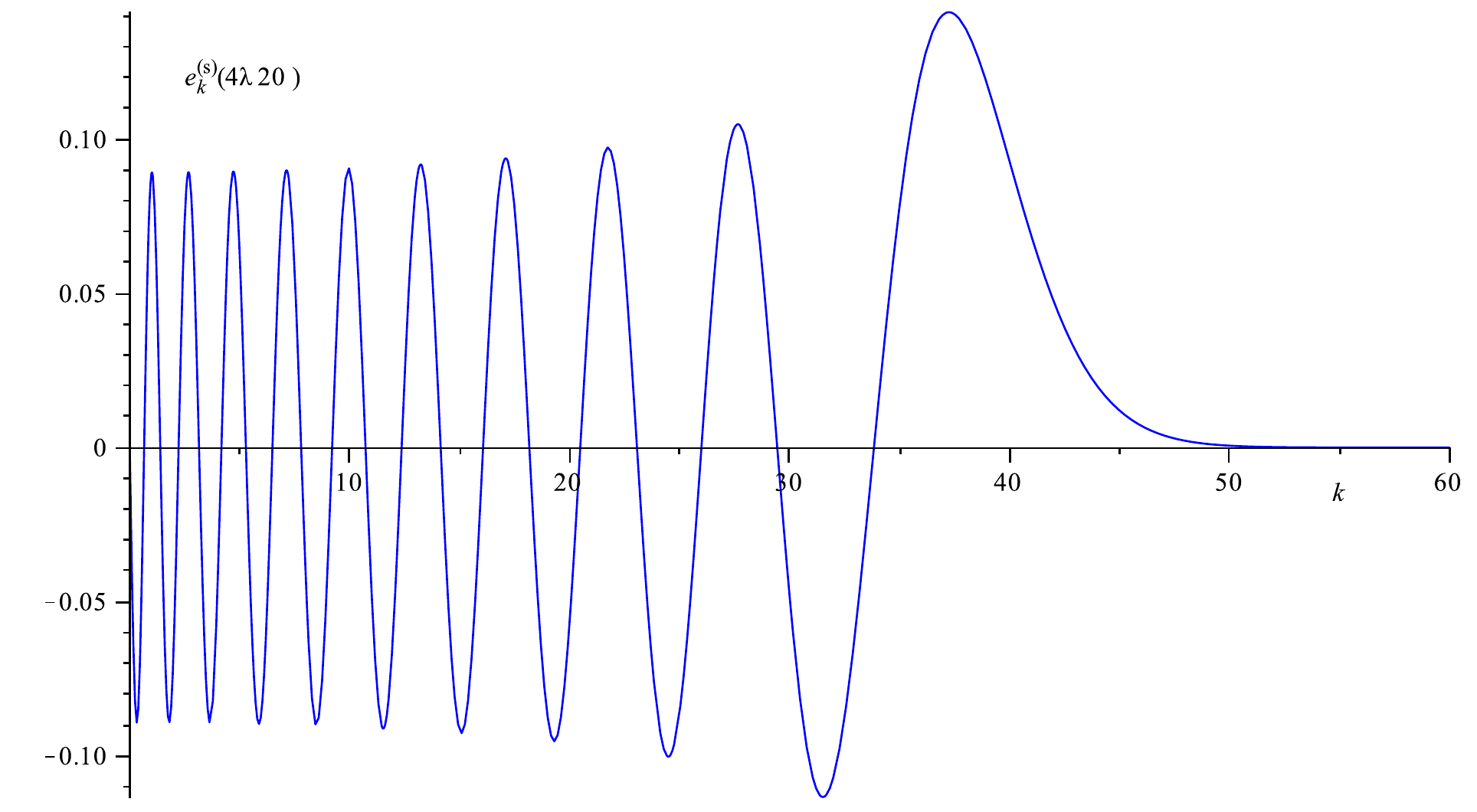}
\label{fig:eskk-a}
}%
\subfloat[$n=200$]{
\includegraphics[width=0.50\textwidth]{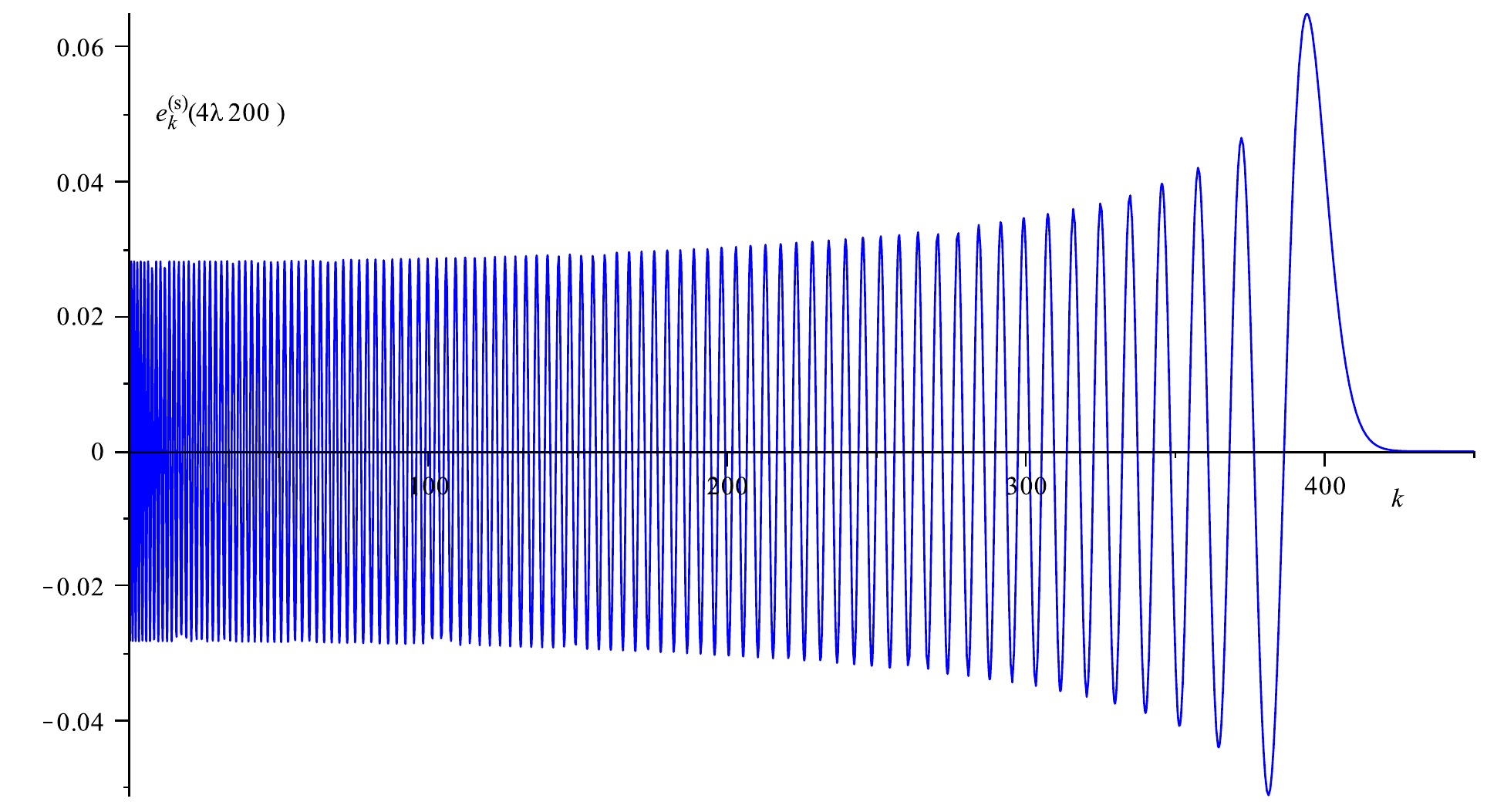}
\label{fig:eskk-b}%
}%
\caption{Plot of $e^{(s)}_k(\v=4\lambda n)$ as a function of $k\geq0$ for
$n=20$ and $n=200$.  
Regarded as a function of $k$, $e^{(s)}_k(\v)$ is symmetric in $k$ and always
exhibits exactly $n-1$ nodes in addition to the node at $k=0$.  
The largest zero and last maximum of $e^{(s)}_k(\v)$ always appears at a 
value $|k_{\textit{max}}| \lesssim 2|n|$, after which $e^{(s)}_k(\v)$ is 
exponentially damped in $k$.  This ultraviolet cutoff at 
$|k|=2|n|=|\v/2\lambda|$ in the eigenfunctions will be explained analytically 
in the sequel.
}%
\label{fig:eskk}%
\end{figure}

\begin{figure}[!htbp]   
\includegraphics[width=1.0\textwidth]{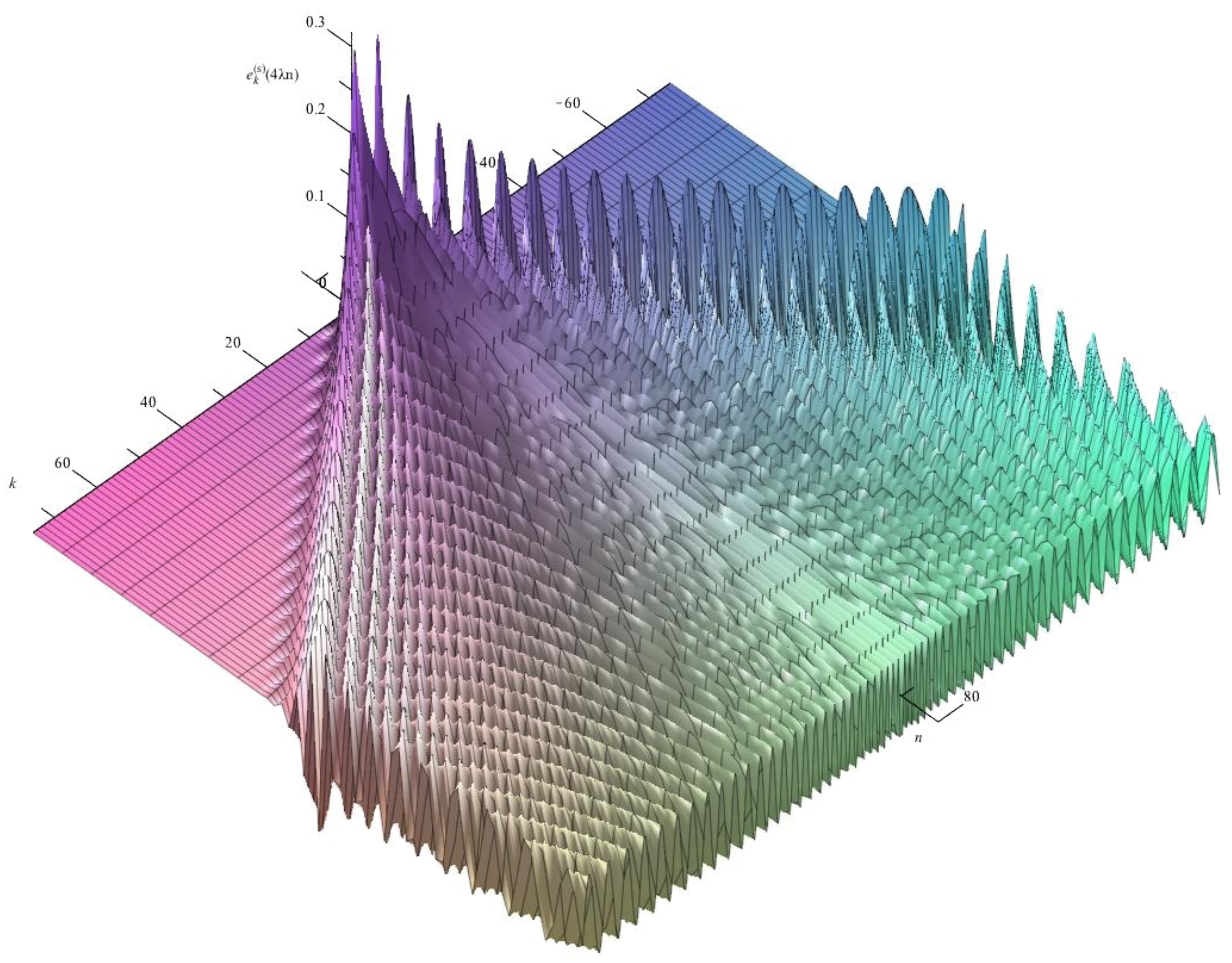}
\caption{Plot of the functions $e^{(s)}_k(\v=4\lambda n)$ in the $(k,n)$ plane
for $0\leq n \leq 75$ and $|k|<75$.  They are symmetric in both $k$ and $n$.
The volume variable $\v=4\lambda n$ is fundamentally discrete; the values of
the eigenfunctions are plotted as continuous in both variables $k$ and $n$ for
reasons of visual clarity only.  (The functions $e^{(s)}_k(\v)$ were evaluated
only at integer values of $n$ to construct this surface, of course.)  The
plots in Fig.\ \ref{fig:eskk} showing the dependence of the $e^{(s)}_k(\v)$ on
$k$ at fixed $n$ may be viewed as constant-$n$ cross-sections of this surface.
Similarly, Fig.\ \ref{fig:eskn}, showing the dependence of the $e^{(s)}_k(\v)$
on $n$ at fixed $k$, may be viewed as constant-$k$ cross-sections of this
surface.  The exponential ultraviolet cutoff along the lines
$|k|=2|n|=|\v/2\lambda|$ is clearly evident.
The dynamical eigenfunctions $e^{(s)}_k(\v)$ may therefore be regarded to an
excellent approximation as having support only in the ``wedge'' $|k|\lesssim
2|n|$.  It is this feature of the eigenfunctions that is ultimately
responsible for the existence of a universal upper bound to the matter density.
}%
\label{fig:esWedge}%
\end{figure}

\begin{figure}[hbtp!]
\subfloat[$k=10$]{
\includegraphics[width=0.45\textwidth]{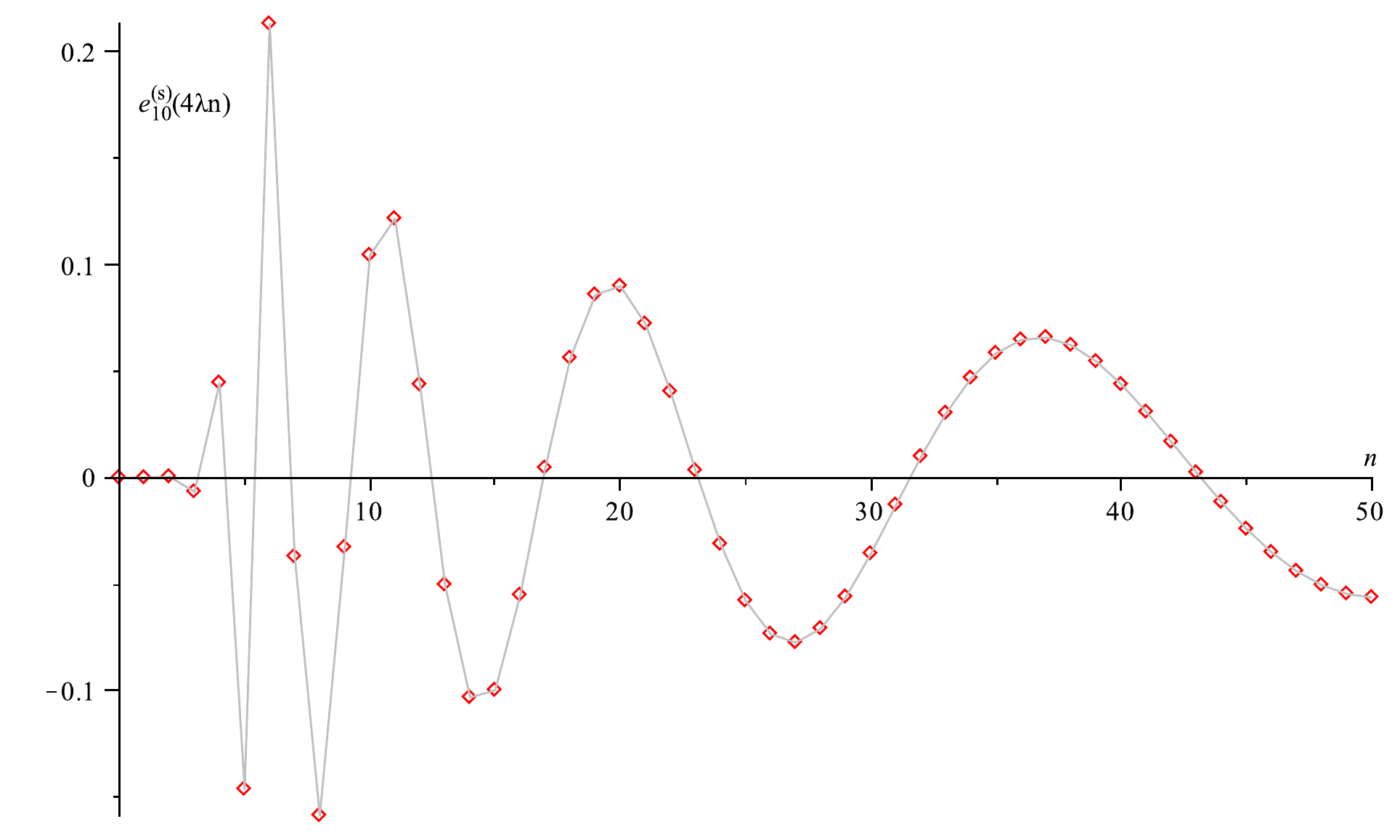}
\label{fig:eskn-a}
}
\subfloat[$k=100$]{
\includegraphics[width=0.45\textwidth]{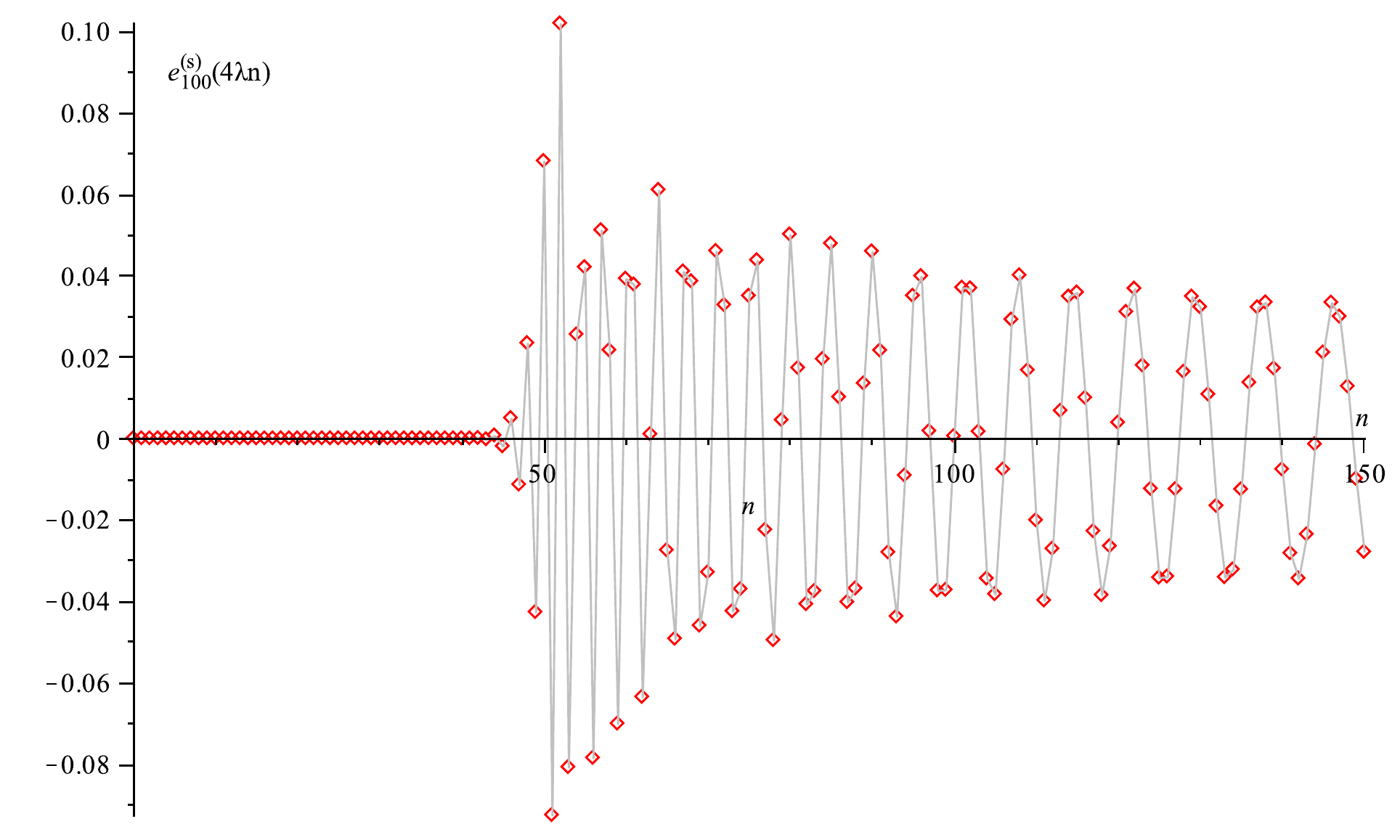}
\label{fig:eskn-b}
} \caption{Plot of $e^{(s)}_k(\v=4\lambda n)$ as a function of $n\geq0$ for
$k=10$ and $k=100$.  Regarded as a function of $n\in\Int$, the $e^{(s)}_k(\v)$
are symmetric in $n$ and have support only on the lattice $\v = 4\lambda n$;
the points are connected for visual clarity only.  Note $e^{(s)}_k(0)=0$ for
$k\neq 0$.  The functions $e^{(s)}_k(\v=4\lambda n)$ for fixed $k$ decay
rapidly to essentially zero for $|n|\lesssim |k|/2$, the ultraviolet cutoff in
the eigenfunctions also visible in Figs.\ \ref{fig:eskn}-\ref{fig:esWedge}.
(The cutoff is not as sharp viewed on slices of constant $k$ as it is on
slices of constant $n$, on which the cutoff is truly exponential.)  For
$|n|\gtrsim |k|/2$, they settle rapidly into a regular decaying oscillation.
This latter behavior corresponds precisely to a symmetric superposition of
Wheeler-DeWitt eigenfunctions to be elaborated in the sequel.
%
}%
\label{fig:eskn}%
\end{figure}

Plots of $e^{(s)}_k(\v)$ are shown in Figs.\ \ref{fig:eskk}, \ref{fig:esWedge}
and \ref{fig:eskn}.  Two behaviors are clearly evident in these plots.  First,
the dynamical eigenfunctions are exponentially damped as functions of $k$ for
$|k| > 2|n| = |\v/2\lambda|$ (Figs.\ \ref{fig:eskk}-\ref{fig:esWedge}).  This
is the ultraviolet momentum space cutoff in the eigenfunctions described in
the introduction.
The cutoff can be understood as a consequence of the fundamental discreteness
of volume in these quantum theories.  Second, for $|n| > |k|$, the
eigenfunctions $e^{(s)}_k(\v)$ settle quickly into a decaying sinusoidal
oscillation in $n$ (Fig.\ \ref{fig:eskn}).  We will see that this oscillation
corresponds to a specific symmetric superposition of the eigenfunctions for
the Wheeler-DeWitt quantization of the same physical model.  As a consequence,
generic quantum states in this loop quantum cosmology will evolve to a
symmetric superposition of an expanding and a collapsing Wheeler-DeWitt
universe.

\subsection{Asymptotics}
\label{sec:eAsymptotics}

It is evident from Fig.\ \ref{fig:esWedge} that, to an excellent 
approximation, the dynamical eigenfunctions $e^{(s)}_k(\v)$ may be regarded 
as having support only in the wedge $|k| \lesssim 2|n|$ in the $(k,n)$ 
plane.  We now seek to explain this behavior based on an analysis of the 
exact solutions, Eq. (\ref{eq:eskdef}), as well as the large volume 
limit and other features visible in Figs.\ \ref{fig:eskk}-\ref{fig:eskn}.  

From Eqs.\ (\ref{eq:eskdef}) and (\ref{eq:Ikn}), $e^{(s)}_k(\v)$ is given by
$\sqrt{|n|/2|k\cdot\sinh(\pi k)|}$ times the polynomial $I(k,|n|)$.  Indeed,
examination of $I(k,|n|)$, regarded as a polynomial in $k$, shows that it is
an even polynomial in $k$ with no constant term, whose terms alternate in
sign.  Thus we see immediately that $e^{(s)}_k(0)=0$ and $e^{(s)}_0(\v)=0$ for
$\v\neq 0$ (cf.\ Eq.\ (\ref{eq:e0-a})).  Examination of plots of $I(k,|n|)$
(as in Fig.\ \ref{fig:eskk}) shows that it always exhibits the maximum number
of roots possible $(2n-1)$ for such a polynomial; the alternating signs of the
coefficients lead to the oscillations.

Writing the product in Eq.\ (\ref{eq:Ikn-a}) as a ``falling factorial'' 
\cite{AS64}, it is possible to arrive at an explicit expression for 
$I(k,|n|)$ which is useful for some computations:%
\footnote{One must be careful, however.  The alternating signs of the
coefficients and the large powers of $k$ appearing in the expression for
$I(k,|n|)$ at large volume ($n$) can quickly lead to numerical instabilities.
} %
\begin{equation}
I(k,|n|) = \sum_{j=1}^{|n|} a(|n|,j)\, k^{2j},
\label{eq:Iknseries}
\end{equation}
where the coefficients $a(|n|,j)$ are given by
\begin{equation}
a(n,j) = (-1)^j \sum_{l=2j}^{2n} s(2n-1,l-1)\cdot \binom{l-1}{2j-1}\cdot
    \left( \sum_{m=0}^{2n} \frac{(m-1)^{l-2j}}{m!(2n-m)!} \right).
\label{eq:Iknseriescoeffs}
\end{equation}
Here the $s(p,q)$ denote the (signed) Stirling numbers of the first kind.%
\footnote{The rising and falling factorials are defined by $x^{\overline{n}} = 
\prod_{i=0}^{n-1}(x+i)$ and $x^{\underline{n}} = \prod_{i=0}^{n-1}(x-i)$.  
The signed Stirling numbers of the first kind are then defined by 
$x^{\underline{n}}=\sum_{i=0}^{n} s(n,i)\,x^i$.
} %
It should be noted that the factor to the right of $(-1)^j$ is always
positive, leading to the alternating signs of these coefficients.

For large $|k|$, $I(k,|n|)$ is dominated by $k^{2|n|}$, and therefore 
\begin{subequations}
\begin{eqnarray}
e^{(s)}_k(\v=4\lambda n) & \sim & \frac{1}{\sqrt{k\sinh(\pi k)}} \cdot k^{2|n|}
\label{eq:eskdecay-a}\\
 & \sim & |k|^{2n-\frac{1}{2}}\cdot e^{-\pi k/2},
\label{eq:eskdecay-b}
\end{eqnarray}
\label{eq:eskdecay}%
\end{subequations}
and the decay of the eigenfunctions is indeed exponential in $k$ past the 
largest root of $e^{(s)}_k(\v)$.

\subsubsection{Steepest descents}
\label{sec:descents}

To identify the value of $k$ at which the decay of the symmetric
eigenfunctions $e^{(s)}_k(\v)$ sets in requires a bit more work.  Recall from
Eq.\ (\ref{eq:eskek}) that the $e^{(s)}_k(\v)$ may be expressed in terms of
the primitive eigenfunctions $e_k(\v)$ given by Eq.\ (\ref{eq:ek}).  The
integral in this equation is of the form
\begin{equation}
\scri(k,\v) = \int_{0}^{\frac{\pi}{\lambda}} db\ e^{if(b,k,\v)},
\label{eq:Ikv}
\end{equation}
where
\begin{equation}
f(b,k,\v) = k\cdot\ln(\tan\frac{\lambda b}{2}) -\frac{\v b}{2}.
\label{eq:fdef}
\end{equation}
This was the form from which the exact expression Eq.\ (\ref{eq:epmkexplicit})
was extracted in Ref.\ \cite{ach10a}.  We however wish to evaluate this
integral in the limits of large $|k|$ and $|\v|$.  While Eq.\ (\ref{eq:Ikv}) 
is not quite of the same form for which the steepest descents approximation 
is normally discussed -- a single large parameter multiplying an overall 
phase -- the same arguments for the validity of the approximation apply.  In 
regions where $f(b,k,\v)$ is large, the integrand oscillates rapidly and 
contributions from neighboring values of $b$ cancel one another.  The 
dominant contributions to $\scri(k,\v)$, therefore, come from regions close 
to the stationary points of $f(b,k,\v)$ where $f$ changes only slowly with 
$b$ and the cancellations are not strong.  This is the usual 
steepest-descents approximation, and in general one has \cite{DK67}
\begin{equation}
\int dz\, e^{if(z)} g(z) \approx 
\sum_{i} \sqrt{\frac{2\pi}{|f''(z_i)|}}\, e^{if(z_i)}\, g(z_i) \,
e^{i\frac{\pi}{4}\sgn(f''(z_i))},
\label{eq:sdapprox}
\end{equation}
where the $z_i$ locate the stationary points $f'(z_i)=0$ along the relevant 
contour.

It is clear that when $|k|$ or $|\v|$ are large, $f(b,k,\v)$ can become 
large, suppressing the value of $\scri(k,\v)$, and so we seek the stationary 
points of $f$.
%

First observe that $f(b,k,\v)$ diverges at $b=0$ and $b=\pi/\lambda$, so there
is no contribution to $\scri(k,\v)$ from the endpoints of the integration due
to the rapid oscillation of the integrand there.  Next, one finds
\begin{equation}
\frac{\partial f}{\partial b}(b,k,\v) = 
    \frac{\lambda k}{\sin\lambda b} - \frac{\v}{2}
\label{eq:fprime}
\end{equation}
and
\begin{equation}
\frac{\partial^2 f}{\partial b^2}(b,k,\v) = 
    -\frac{\lambda^2 k\cos\lambda b}{\sin^2\lambda b}.
\label{eq:f2prime}
\end{equation}
The stationary points therefore satisfy
\begin{equation}
\sin\lambda b = \frac{2\lambda k}{\v}.
\label{eq:fstationary}
\end{equation}
When solutions exist there are two roots $b_1$ and $b_2$, given in the limit 
$|\v| \gg \lambda|k|$ by
\begin{subequations}
\begin{eqnarray}
b_1 & \approx & \frac{2k}{\v},
\label{eq:b-a}\\
b_2 & \approx & \frac{\pi}{\lambda} - \frac{2k}{\v}.
\label{eq:b-b}
\end{eqnarray}
\label{eq:b}%
\end{subequations}
In this limit
\begin{subequations}
\begin{eqnarray}
f(b_1,k,\v) & \approx & -k\left[ \ln\frac{\v}{\lambda k} + 1 \right],
\label{eq:fstat-a}\\
f(b_2,k,\v) & \approx & +k\left[ \ln\frac{\v}{\lambda k} + 1 \right] 
   - \frac{\pi\v}{2\lambda},
\label{eq:fstat-b}
\end{eqnarray}
\label{eq:fstat}%
\end{subequations}
and 
\begin{subequations}
\begin{eqnarray}
\frac{\partial^2 f}{\partial b^2}(b_1,k,\v) & \approx & -\frac{\v^2}{4k},
\label{eq:f2stat-a}\\
\frac{\partial^2 f}{\partial b^2}(b_2,k,\v) & \approx & +\frac{\v^2}{4k}.
\label{eq:f2stat-b}
\end{eqnarray}
\label{eq:f2stat}%
\end{subequations}
We now piece together these results to study the asymptotic limits of 
$\scri(k,\v)$ and consequently $e^{(s)}_k(\v)$.

\subsubsection{Wheeler-DeWitt limit}
\label{sec:eWdW}

We will begin by considering the case in which $k$ and $\v$ are both 
positive, and return to the other possibilities shortly.  When $\v \gg 
\lambda k > 0$, from Eq.\ (\ref{eq:sdapprox}) we find
\begin{subequations}
\begin{eqnarray}
\scri(k,\v) & \approx & 
\sqrt{\frac{2\pi}{|f''(b_1)|}}\, e^{if(b_1)}\, e^{i\frac{\pi}{4}\sgn(f''(b_1))}
+
\sqrt{\frac{2\pi}{|f''(b_2)|}}\, e^{if(b_2)}\, e^{i\frac{\pi}{4}\sgn(f''(b_2))}
\label{eq:Ikvsd-a}\\
 & \cong & \sqrt{\frac{8\pi|k|}{\v^2}}\left\{
 e^{-ik\left[\ln\frac{\v}{\lambda k}+1\right]} e^{-i\frac{\pi}{4}}  +
 e^{+ik\left[\ln\frac{\v}{\lambda k}+1\right]} e^{+i\frac{\pi}{4}} 
 e^{-i\frac{\pi\v}{2\lambda}}   \right\}
\label{eq:Ikvsd-b}\\
 & = & 2 \sqrt{\frac{8\pi|k|}{\v^2}} 
      \cos\left(k\left[\ln\frac{\v}{\lambda  k}+1\right] + \frac{\pi}{4}\right),
\label{eq:Ikvsd-c}
\end{eqnarray}
\label{eq:Ikvsd}%
\end{subequations}
where to get to the last line we recall $\v=4\lambda n$, so the final 
exponential factor in Eq.\ (\ref{eq:Ikvsd-b}) is unity.%

In the case where $k$ and $\v$ are both negative, the same results obtain, 
but now
\begin{subequations}
\begin{eqnarray}
f(b_1,k,\v) & \approx & +|k|\left[ \ln\frac{\v}{\lambda k} + 1 \right]
   \phantom{- \frac{\pi\v}{2\lambda}}
   \qquad\   \sgn(f''(b_1)) = +,
\label{eq:fstatabs-a}\\
f(b_2,k,\v) & \approx & -|k|\left[ \ln\frac{\v}{\lambda k} + 1 \right] 
   - \frac{\pi\v}{2\lambda}
      \qquad   \sgn(f''(b_2)) = -,
\label{eq:fstatabs-b}
\end{eqnarray}
\label{eq:fstatabs}%
\end{subequations}
again leading to a cosine, but with $k\rightarrow|k|$.   Thus, from Eqs.\ 
(\ref{eq:ek}), (\ref{eq:Ak}), and (\ref{eq:Ikvsd-c}), we find%
\footnote{We could also have arrived at the absolute value signs simply by 
observing from Eq.\ (\ref{eq:ekvsvk}) that $e_{-k}(-\v)=e_{k}(\v)$.
} %
\begin{equation}
e_k(\v) \cong \frac{2}{\sqrt{|\sinh(\pi k)}|} \sqrt{\frac{2\lambda}{\pi|\v|}}
  \cos\left(|k|\ln\left|\frac{\v}{\lambda}\right|+\alpha(|k|)\right)
   \text{ when }
  \left\{
  \begin{array}{lcl}   
   |\v|\gg\lambda|k|     \\
   \v\cdot k > 0
  \end{array}\right. , 
\label{eq:ekvWdW}
\end{equation}
where
\begin{equation}
\alpha(k) = k(1-\ln k) + \frac{\pi}{4}.
\label{eq:alphak}
\end{equation}

We have yet to consider the case where $k$ and $\v$ are opposite in sign.  In 
this case note that since $0\leq b \leq \pi/\lambda$, there \emph{are no 
solutions} to Eq.\ (\ref{eq:fstationary}), and $f(b,k,\v)$ has no stationary 
points in the domain of integration.  Thus $\scri(k,\v)$, and hence 
$e_k(\v)$, are strongly suppressed by the rapid oscillations of the integrand 
when $k$ and $\v$ are opposite in sign. 

We note from Eq.\ (\ref{eq:eskek}) that the symmetric eigenfunctions
$e^{(s)}_k(\v)$ are a linear combination of $e_{k}(\v)$ and $e_{-k}(\v)$.  The
functions $e_{-k}(\v)$ are, \emph{mutatis mutandis} as above, strongly
suppressed when $\v$ and $k$ have the \emph{same} sign, and assume the limit 
Eq.\ (\ref{eq:ekvWdW}) when $k$ and $\v$ are opposite in sign.  The 
$|\v|\gg\lambda |k|$ limit of $e^{(s)}_k(\v)$ will therefore pick up precisely 
one contribution of the form of Eq.\ (\ref{eq:ekvWdW}) no matter the signs of 
$k$ and $\v$.  Observing that $\cosh(\pi k/2)/\sqrt{\sinh(\pi|k|)}\approx 
1/\sqrt{2}$ for even very modest values of $k\gtrsim 1$, we arrive finally at
\begin{equation}
e^{(s)}_k(\v) \cong \sqrt{\frac{2\lambda}{\pi|\v|}}
  \cos\left(|k|\ln\left|\frac{\v}{\lambda}\right|+\alpha(|k|)\right)
  \qquad |\v|\gg\lambda |k|.
\label{eq:eskWdW}
\end{equation}

\begin{figure}[hbtp]%
\includegraphics[width=0.95\textwidth]{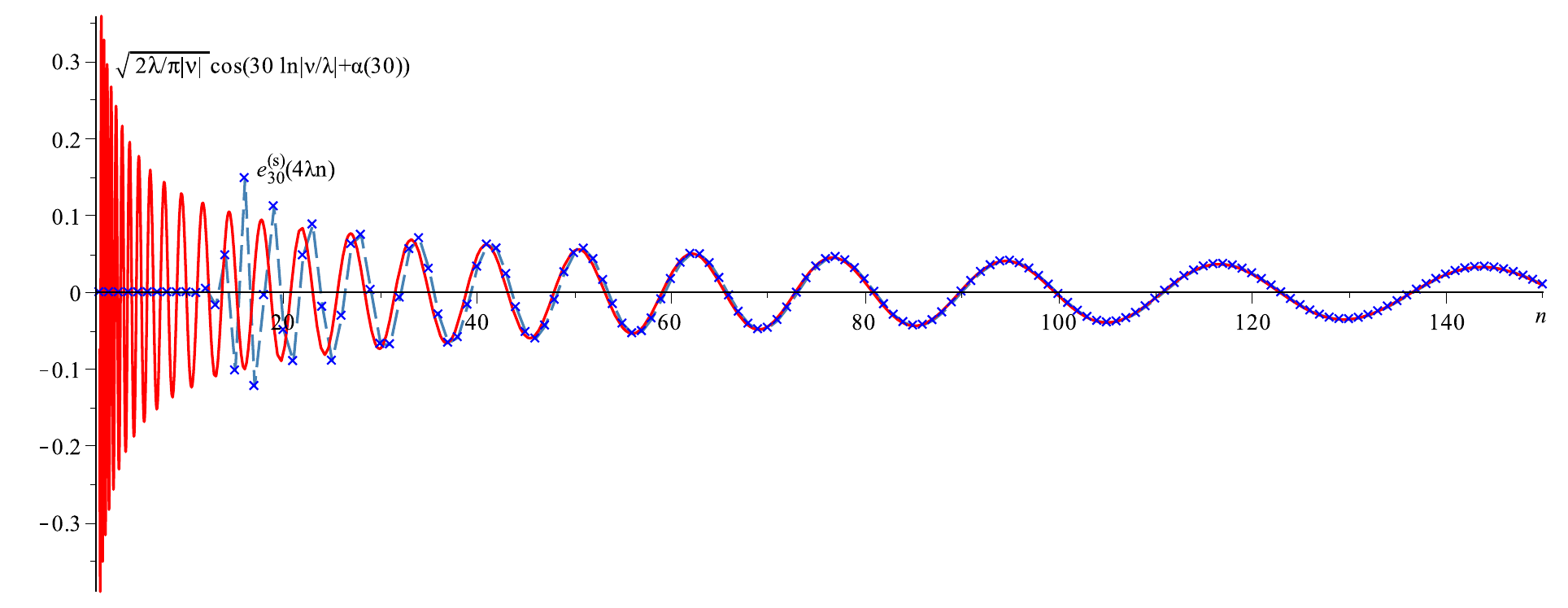}%
\caption{Plot as a function of $n$ of both the dynamical eigenstate
$e^{(s)}_k(\v=4\lambda n)$ and the asymptotic form Eq.\ (\ref{eq:eskWdW}) for
$k=30$.  The volume variable $\v=4\lambda n$ is fundamentally discrete; the
values of $e^{(s)}_k(n)$ are marked with blue ${\scriptstyle\times}$'s; the
points are connected by a dashed blue line for visual clarity.  The solid red
curve is the corresponding asymptotic form.
The rapid convergence to the asymptotic form for $|n|\gg |k|$ on the lattice
$\v=4\lambda n$ is clear.  Note this asymptotic form corresponds to the
particular superposition of eigenstates of the Wheeler-DeWitt quantization of
the same model given by Eq.\ (\ref{eq:eskWdWz}).  The rapid oscillations
visible at small volume are the correct physical behavior of the
Wheeler-DeWitt states, and are ultimately responsible for the fact that these
models are singular in the Wheeer-DeWitt quantization.  See Refs.\
\cite{CS10c,CS12b} for further discussion.
}%
\label{fig:esWdWlimit}%
\end{figure}

Fig.\ \ref{fig:esWdWlimit} shows that the exact eigenfunctions settle down to
this asymptotic form very quickly.  We will employ Eq.\ (\ref{eq:eskWdW}) to
study in Sec.\ \ref{sec:largeV} the large volume limit of flat scalar loop
quantum universes.

The asymptotic expression Eq.\ (\ref{eq:eskWdW}) was, in effect, arrived at on
the basis of analytical and numerical considerations in Ref.\
\cite{aps:improved}, with a numerically motivated fit for the phase
$\alpha(k)$.%
\footnote{Consult footnote \ref{foot:statenorm} concerning the factor 
$1/\sqrt{|\v|}$ leading to the decay of the eigenfunctions with increasing 
volume.
} %
In Ref.\ \cite{kp10a} an expression equivalent to Eq.\ (\ref{eq:eskWdW}) was
derived from a careful analysis of the asymptotic limit of solutions to the
constraint equation, including an expression for the phase $\alpha(k)$
equivalent to Eq.\ (\ref{eq:alphak}).  (See Ref.\ \cite{mbmmo09a} for a
related analysis of this limit.)  Here we have instead derived this
asymptotic form from the exact eigenfunctions, explicitly confirming these
prior analyses with the exact solutions for the model.

\subsubsection{Ultraviolet Cutoff}
\label{sec:eUV}

Figures \ref{fig:eskk}-\ref{fig:esWedge} clearly exhibit the exponential 
ultraviolet cutoff in the eigenfunctions $e^{(s)}_k(\v)$ for values of 
$|k|>2|n| = |\v/2\lambda|$.  We know already from Eq.\ (\ref{eq:eskdecay}) 
that an exponential decay will eventually set in.  The only question is, at 
what value of $k$ does that occur?   We have, in fact, already seen the 
origin of this cutoff and its value.  Eq.\ (\ref{eq:fstationary}) shows that 
$f(b,k,|\v|)$ has no stationary points when $|2\lambda k/\v| > 1$.  In other 
words, $\scri(k,\v)$, hence $e_{k}(\v)$ and $e^{(s)}_k(\v)$, are strongly 
suppressed unless
\begin{subequations}
\begin{eqnarray}
|k| & \lesssim & \left|\frac{\v}{2\lambda}\right|
\label{eq:UVcutoff-a}\\
 & = & 2|n|.
\label{eq:UVcutoff-b}
\end{eqnarray}
\label{eq:UVcutoff}%
\end{subequations}
This cutoff -- in particular, its linear scaling with volume -- may be
understood physically as a consequence of the underlying discreteness of the
quantum geometry.  States with wave numbers $|k|>2|n|$ (i.e.\ wavelengths
shorter than the scale set by $|\lambda/\v|$) are not supported.  
Alternately, it may be viewed as the manifestation in the eigenfunctions of 
the ``quantum repulsion'' generated by quantum geometry at volumes smaller 
than the wave number.


\subsubsection{Small volume limit}
\label{sec:smallV}

The same argument shows that the eigenstates will, equivalently, decay rapidly 
for small volume, when $ |n| \lesssim |k|/2$, as is clear in Fig.\ 
\ref{fig:eskn}.  Eq.\ (\ref{eq:eskdecay}) tells us the 
decay in $e^{(s)}_k(\v)$ as a function of $k$ is exponential.  The precise 
functional form of the decay as a function of $n$ is less evident, but the 
figures show it is also quick.

At this point a comment may be in order.  It is tempting to study this
question by regarding $e^{(s)}_k(\v)$ as a function of a \emph{continuous}
variable $\v$.  However, plotting the exact expressions for $e^{(s)}_k(\v)$
for continuous values of $\v$ on top of the values for $\v=4\lambda n$ should
quickly disabuse one of the notion that there is a simple sense in which
$e^{(s)}_k(\v)$ is well approximated by its naive continuation to the
continuum.  In fact, as discussed in detail in Ref.\ \cite{acs:slqc}, the
convergence to the Wheeler-DeWitt theory in the continuum is not uniform, and
must be extracted with some care in the limit the ``area gap'' set by
$\lambda$ -- fixed in loop quantum cosmology to the value of Eq.\
(\ref{eq:lambdadelta}) -- tends to 0.

As noted in the Introduction, and as is clearly evident in Fig.\
\ref{fig:esWedge}, the ultraviolet cutoff in momentum space is the ``same''
cutoff as the rapid decay at small volume as a function of volume that has
long been known in loop quantum cosmology based on numerical solutions for the
eigenfunctions \cite{aps,aps:improved}.  It was also known numerically that
the onset of this decay was proportional to the eigenvalue $\omega_k$.  (See
e.g.\ Refs.\ \cite{apsv07a,bp08a}.)  What is new in the present work,
facilitated by the change in perspective to consideration of the behavior of
the eigenfunctions as functions of the continuous variable $k$, is an analytic
understanding of the linear cutoff grounded in a study of the model's exact
solutions, its precise value, and its specific relation to the critical
density.

\subsection{Representation of operators}
\label{sec:repop}

As noted above in Eq.\ (\ref{eq:volsym}), we have restricted attention to the 
volume-symmetric sector of the theory.  This is only possible because the physical
operators preserve the symmetry of the quantum states.  

From Eq.\ (\ref{eq:theta}), the matrix elements of $\Theta$ in the volume basis 
may be expressed as
\begin{equation}
\melt{\v}{\Theta}{\v'} = 12\pi G \sqrt{|n\cdot n'|}|n+n'|\cdot
 \left\{ \delta_{n,n'}-\frac{1}{2}\left[\delta_{n,n'+1}+\delta_{n,n'-1}  \right]\right\},
\label{eq:thetameltv}%
\end{equation}
where $\v=4\lambda n$ and $\v'=4\lambda n'$.  Note these matrix elements satisfy 
the following properties:
\begin{subequations}
\begin{eqnarray}
\melt{\v}{\Theta}{\v} & = & \melt{-\v}{\Theta}{-\v}
\label{eq:thetasymv-a}\\
\melt{\v}{\Theta}{\v'} & = & \melt{\v}{\Theta}{\v'}^*
\label{eq:thetasymv-b}\\
                       & = & \melt{\v'}{\Theta}{\v}.
\label{eq:thetasymv-c}%
\end{eqnarray}
\label{eq:thetasymv}%
\end{subequations}
Owing to these relations, the operator $\Theta$ preserves the subspaces
$\Hphys^{(s)}$ and $\Hphys^{(a)}$ of states that are even and odd in $\v$, so
that $P^{(s)}\Theta P^{(a)}=0$, where $P^{(s)}$ and $P^{(a)}$ are the
corresponding projections.  (In other words, $\Theta$ commutes with the parity
operator $\Pi_{\v}=P^{(s)}-P^{(a)}$ \cite{aps,aps:improved}.)  On the
symmetric subspace $\Hphys^{(s)}$ to which we have restricted ourselves,
$\Theta|_{\Hphys^{(s)}} = P^{(s)}\Theta P^{(s)}\equiv \Theta^{(s)}$ (and
correspondingly $p_{\phi}^{(s)}=\hbar\sqrt{\Theta^{(s)}}$) may be decomposed
in terms of the symmetric basis of eigenstates $\ket{k^{(s)}}$,
\begin{equation}
\Theta^{(s)} = \k^2\int dk\, k^2 \ketbra{k^{(s)}}{k^{(s)}},
\label{eq:thetasdecomp}
\end{equation}
where $e^{(s)}_{k}(\v) \equiv \bracket{\v}{k^{(s)}}$.
The matrix elements of $\Theta^{(s)}$ in the volume representation are related 
to those of $\Theta$ by
\begin{subequations}
\begin{eqnarray}
\melt{\v}{\Theta^{(s)}}{\v'} & = & \frac{1}{2}
      \left\{\melt{\v}{\Theta}{\v'} + \melt{\v}{\Theta}{-\v'}  \right\}
\label{eq:thetastheta-a}\\
 & = & \melt{\v}{\Theta^{(s)}}{-\v'}.
\label{eq:thetastheta-b}%
\end{eqnarray}
\label{eq:thetastheta}%
\end{subequations}
The actions of $\Theta$ and $\Theta^{(s)}$ on $\Hphys^{(s)}$ are of course 
completely equivalent.  Since $\hat{p}_{\phi}^2 = \hbar^2\Theta$ on $\Hphys$, these
expressions give the matrix elements of $\hat{p}_{\phi}^2$ on $\Hphys^{(s)}$ as well.

The matrix elements of $\hat{p}_{\phi}=\hbar\sqrt{\Theta}$ are more complex.
These are given in terms of derivatives of a generating function in Appendix C
of Ref.\ \cite{ach10a}.  Explicit expressions for the physical observables in
another representation are also given in Ref.\ \cite{acs:slqc}.  Here we note
that on $\Hphys^{(s)}$ we may employ the $e^{(s)}_{k}(\v)$ to calculate
$\hat{p}^{(s)}_{\phi}$ explicitly in the volume representation.  Indeed, the
polynomial solution Eq.\ (\ref{eq:Iknseries}) for $I(k,n)$ makes it a
straightforward matter to evaluate these matrix elements.  The result is (with
$\v=4\lambda n$ and $\v'=4\lambda m$)
\ifthenelse{\pp=1}{%
\begin{multline}
\int_{-\infty}^{\infty}\!dk\, |k|\, e^{(s)}_k(\v) e^{(s)}_k(\v')^* =\\ 
\sqrt{|n\cdot m|}\, \sum_{j=1}^{|n|}\sum_{l=1}^{|m|} a(|n|,j) a(|m|,l)
\frac{2^{2(j+l)+1}-1}{2^{2(j+l)}\pi^{2(j+l)+1}}\Gamma(2(j+l)+1)\zeta(2(j+l)+1)\\
\approx 
\sqrt{|n\cdot m|}\, \sum_{j=1}^{|n|}\sum_{l=1}^{|m|} a(|n|,j) a(|m|,l)
\frac{2}{\pi^{2(j+l)+1}}\Gamma(2(j+l)+1),
\label{eq:pphimeltsoln}
\end{multline}
}{%
\begin{subequations}
\begin{eqnarray}
\int_{-\infty}^{\infty}\!dk\, |k|\, e^{(s)}_k(\v) e^{(s)}_k(\v')^* & = & 
\sqrt{|n\cdot m|}\, \sum_{j=1}^{|n|}\sum_{l=1}^{|m|} a(|n|,j) a(|m|,l)
\frac{2^{2(j+l)+1}-1}{2^{2(j+l)}\pi^{2(j+l)+1}}\Gamma(2(j+l)+1)\zeta(2(j+l)+1)
\nonumber\\
\label{eq:pphimeltsoln-a}\\
 & \approx & 
\sqrt{|n\cdot m|}\, \sum_{j=1}^{|n|}\sum_{l=1}^{|m|} a(|n|,j) a(|m|,l)
\frac{2}{\pi^{2(j+l)+1}}\Gamma(2(j+l)+1),
\label{eq:pphimeltsoln-b}%
\end{eqnarray}
\label{eq:pphimeltsoln}%
\end{subequations}
}%
where $\zeta(z)$ is the Riemann zeta-function.  This expression gives the
$(\v,\v')$ matrix elements of $\hat{p}^{(s)}_{\phi}/\hbar\k$, or equivalently
$\smash{\sqrt{\Theta^{(s)}}}/\k$.

\begin{figure}[hbtp!]
\includegraphics[width=0.75\textwidth]{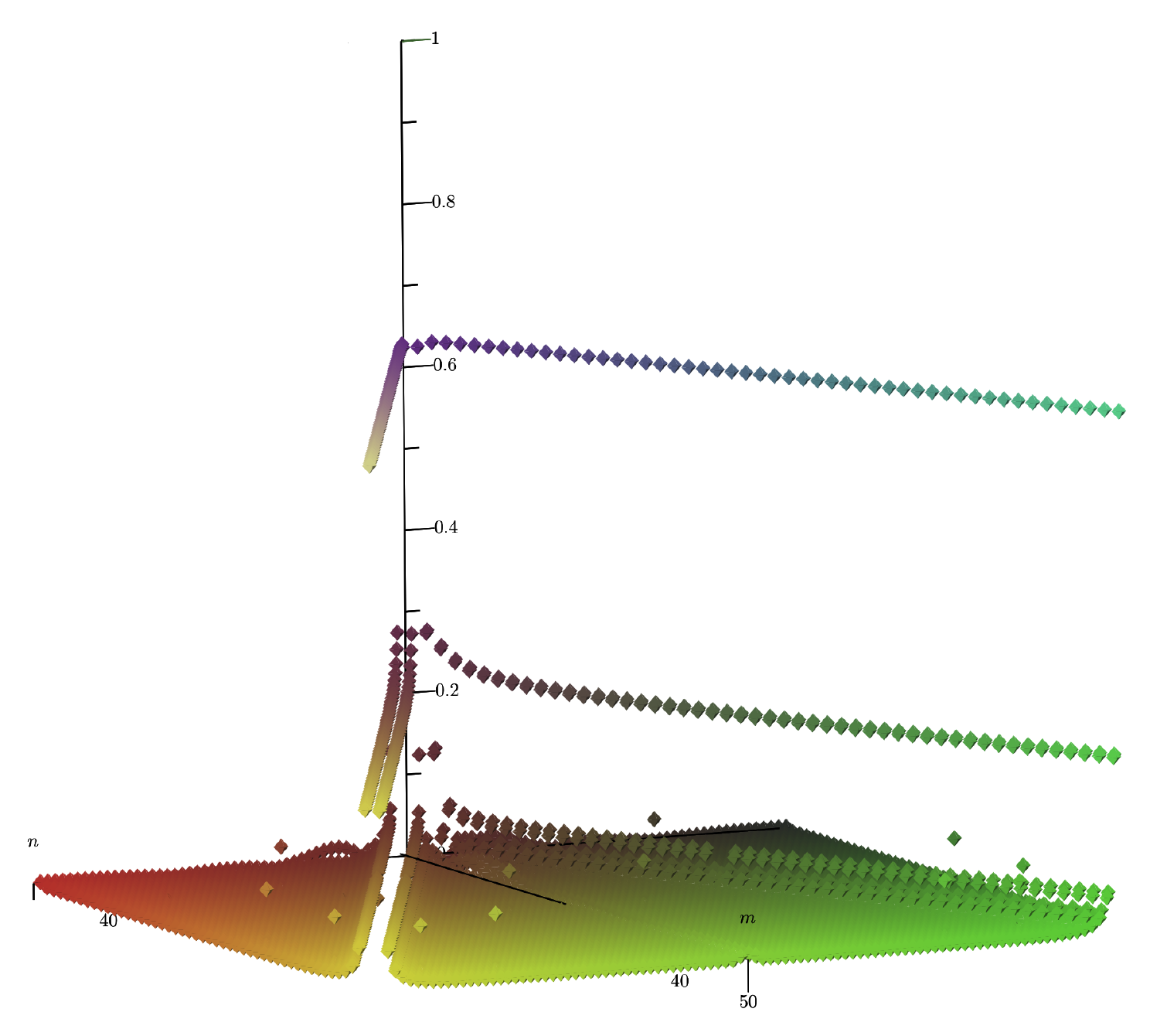}
\caption{Plot of the norm of the scalar momentum matrix
element overlap integral appearing in Eq.\ (\ref{eq:pphimeltbnd})
normalized by the estimated maximum value $\tfrac{1}{2}|\v/2\lambda|=|n|$ of
this integral on the diagonal $|\v|=|\v'|$, taking $|\v=4\lambda
n|\leq|\v'=4\lambda m|$: $|\int_{-\infty}^{\infty}dk
  |k|e^{(s)}_k(\v)e^{(s)}_k(\v')^*|/|n|$.   %
The integrals have been calculated numerically from the exact eigenfunctions
over the range $0<|n|\leq 50$ and $0<m\leq 50$.  According to the upper bound
expressed in Eq.\ (\ref{eq:pphimeltbnddiagexact}), this normalized matrix
element is bounded above by one, and as argued is strongly suppressed off the
diagonal $|\v|=|\v'|$.  In effect, these plots show that $\hat{p}_{\phi}$
\emph{approximately} commutes with $|\hat{\v}|$ since it is nearly diagonal in
the $\v$-representation.  This is essentially the reason the ``moral''
argument expressed in Eq.\ (\ref{eq:rhoexpctbndmoral}) 
for the existence of a critical density in this model yields the correct
result.
}%
\label{fig:momover}%
\end{figure}

We observe from Eq.\ (\ref{eq:thetameltv}) that $\Theta$ is nearly
diagonal in the volume representation, with only the $n'=n, n\pm 1$ elements
not exactly zero.  The same is therefore true of $\hat{p}_{\phi}^2$.  Direct
numerical evaluation of the expression Eq.\ (\ref{eq:pphimeltsoln}) reveals
that $\hat{p}_{\phi}$ -- and therefore $\sqrt{\Theta}$ -- are also nearly
diagonal in the volume representation, with only the $n'=n, n\pm 1$ matrix
elements significantly different from zero.  (See Fig.\ \ref{fig:momover}.)  In this
case, however, the off-diagonal elements of $\hat{p}_{\phi}$ are merely very
small, rather than precisely zero.

The values of these matrix elements can be understood as a consequence of the 
ultraviolet cutoff, Eq.\ (\ref{eq:UVcutoff}).  Indeed, the exponential 
cutoff $|k| \lesssim |\v/2\lambda|$ implies that the diagonal matrix elements are 
bounded,%
\footnote{Numerical integration of the exact eigenfunctions shows that this
rough estimate of the upper bound typically overestimates the actual value of
the integral by around $60\%$ due to the (suppressed) contribution to the
normalization integral from the increased amplitude near $|k|\sim
|\v'/2\lambda|$.  See Fig.\ \ref{fig:momover}.
} %
\begin{equation}
\int_{-\infty}^{\infty}\!dk\, |k|\, e^{(s)}_k(\v) e^{(s)}_k(\pm\v)^*
 \lesssim \frac{1}{2} \left| \frac{\v}{2\lambda}\right|.
\label{eq:pphimeltbnddiag}
\end{equation}
(The $1/2$ is a consequence of the symmetric normalization of the
eigenfunctions, Eq.  (\ref{eq:eskcomplete}).)  This bound on
$\smash{\sqrt{\Theta^{(s)}}}/\k$ may be compared with that set by the exact
expression for $\Theta$, Eq.\ (\ref{eq:thetameltv}).  From Eq.\
(\ref{eq:thetastheta}),
\begin{subequations}
\begin{eqnarray}
\melt{\v}{\Theta^{(s)}}{\v} & = & \frac{1}{2}\melt{\v}{\Theta}{\v}
\label{eq:thetasthetav-a}\\
 & = & \frac{\k^2}{4}\left|\frac{\v}{2\lambda}\right|^2.
\label{eq:thetasthetav-b}%
\end{eqnarray}
\label{eq:thetasthetav}%
\end{subequations}
As it is always the case that $\expct{\hat{A}^2} \geq \expct{\hat{A}}^2$, we
see that a strict bound on the diagonal matrix elements of
$\smash{\sqrt{\Theta^{(s)}}}/\k$ is
\begin{equation}
\int_{-\infty}^{\infty}\!dk\, |k|\, e^{(s)}_k(\v) e^{(s)}_k(\pm\v)^*
 \leq \frac{1}{2} \left| \frac{\v}{2\lambda}\right|,
\label{eq:pphimeltbnddiagexact}
\end{equation}
in agreement with the bound inferred from the UV cutoff.   

The off-diagonal elements may be bounded in a similar manner.  For simplicity
assume $|\v|<|\v'|$.  The exponential UV cutoff effectively restricts the
range of integration to $|k|\lesssim |\v/2\lambda|$.  The Cauchy-Schwarz
inequality then gives
\begin{equation}
\left|\int_{-\infty}^{\infty}\!dk\, |k| e^{(s)}_k(\v) e^{(s)}_k(\v')^*\right|
\lesssim  \left|  \frac{\v}{2\lambda}\right|
\sqrt{\int_{-|\v/2\lambda|}^{|\v/2\lambda|}\!dk\, |e^{(s)}_k(\v)|^2}
\sqrt{\int_{-|\v/2\lambda|}^{|\v/2\lambda|}\!dk\, |e^{(s)}_k(\v')|^2}.
\label{eq:pphimeltbnd}
\end{equation}
Again, because the $e^{(s)}_k(\v)$ are symmetrically normalized, 
the value of the first square root is essentially
$1/\sqrt{2}$.  As for the second, we note from Fig.\ \ref{fig:eskk} that the
$e^{(s)}_k(\v)$ execute approximately uniform amplitude oscillations, growing
slowly with increasing $k$ with a short lived increase before the exponential
cutoff sets in at $|k|=|\v/2\lambda|$.  Therefore, for $|\v| < |\v'|$ we may
estimate that at most
\begin{equation}
\int_{-|\v/2\lambda|}^{|\v/2\lambda|}\!dk\, |e^{(s)}_k(\v')|^2
 \lesssim \frac{1}{2} \left|\frac{\v}{\v'}\right|,
\label{eq:eskvpcut}
\end{equation}
showing that the cutoff alone implies that the off-diagonal terms are
suppressed relative to the diagonal terms.  Interference effects only reduce
their values further; Fig.\ \ref{fig:momover} shows that except for the $n =
n'\pm 1$ elements -- as with $\Theta$ itself -- this suppression is dramatic.

As a shorthand to express these bounds, we can say that $\hat{p}_{\phi}$ and
$|\hat{\v}|$ -- and therefore $|\hat{\v}|_{\phi}$ -- \emph{approximately}
commute, in the sense that
$\melt{\v}{\hat{p}_{\phi}}{\v'}$ is approximately diagonal.  %
(See Fig.\ \ref{fig:momover}.)  We will see in Sec.\ \ref{sec:rhocrit} that
this helps explain why the matter density in these models remains bounded even
though the spectrum of the scalar momentum $\hat{p}_{\phi}$ is not itself
bounded.

\section{Large volume limit of loop quantum states}
\label{sec:largeV}

In Eq.\ (\ref{eq:eskWdW}) we have exhibited the large volume (more precisely,
$|\v|\gg\lambda|k|$) limit of the basis $e^{(s)}_k(\v)$ of symmetric states of
flat scalar loop quantum cosmology.   
We extracted this limit from the exact solution for the model's
eigenfunctions, essentially confirming prior numerical and analytical work.
In this section we relate these states to the eigenstates in a Wheeler-DeWitt
quantization of the same physical model.

A complete, rigorous Hilbert space quantization of a flat
Friedmann-Lema\^{i}tre-Robertson-Walker cosmology sourced by a massless
minimally coupled scalar field has been given in Refs.\
\cite{aps,aps:improved} and compared to its loop quantization in detail in
Ref.\ \cite{acs:slqc}.  It is known rigorously that states in the
Wheeler-DeWitt quantization are generically singular just as they are in the
classical theory in the sense that all states assume arbitrarily small volume
(equivalently, large density) at some point in their cosmic evolution in
``internal time'' $\phi$ \cite{acs:slqc,CS10c}.

The classical solutions are given in Eq.\ (\ref{eq:classtraj}), corresponding 
to disjoint expanding and contracting branches which either begin or end in 
the classical singularity at $V=0$.   Solutions to the Wheeler-DeWitt quantum 
theory similarly divide into disjoint expanding and contracting branches, and 
as noted, are singular in the same way.

The Wheeler-DeWitt version of the quantum constraint is%
\footnote{See footnote \ref{foot:statenorm}.
} %
\cite{aps:improved}
\begin{subequations}
\begin{eqnarray}
\partial_{\phi}^2 \PsiW(\v,\phi) & = & 
  12\pi G \frac{1}{\sqrt{|\v|}} 
  \   \v\partial_{\v}(\v\partial_{\v}\sqrt{|\v|}\PsiW(\v,\phi))
\label{eq:thetaWdW-a}\\
 & := & \ThetaW_{\v}\PsiW(\v,\phi).
\label{eq:thetaWdW-b}
\end{eqnarray}
\label{eq:thetaWdW}%
\end{subequations}
Attention may again be restricted to symmetric (Eq.\ (\ref{eq:volsym})), 
positive frequency solutions in the sense of Eq.\ (\ref{eq:posfreq}).  The 
symmetric eigenstates of $\ThetaW_{\v}$ satisfying Eqs.\ 
(\ref{eq:thetaefn})-(\ref{eq:omegak}) are
\begin{equation}
\eW_k(\v) = \frac{1}{\sqrt{4\pi|\v|}} e^{ik\ln\left|\frac{\v}{\lambda}\right|},
\label{eq:eWdW}
\end{equation}
and are orthonormal (distributionally normalized to $\delta(k,k')$) in the
inner product
\begin{equation}
\bracket{\PsiW}{\PhiW} = \int_{-\infty}^{\infty}d\v\, \PsiW(\v,\phi)^*\PhiW(\v,\phi)
\label{eq:ipWdW}
\end{equation}
resulting from group averaging.  Physical states may then be expressed as
\begin{subequations}
\begin{eqnarray}
\PsiW(\v,\phi) & = & 
  \int_{-\infty}^{+\infty}dk\, \PsiWk(k)\, \eW_k(\v)\, e^{i\omega_k\phi}
\label{eq:PsiexpWdW-a}\\
 & = & 
  \frac{1}{\sqrt{4\pi|\v|}} \int_{-\infty}^{0}dk\, \PsiWk(k) \,
    e^{ik\left[\ln\left|\frac{\v}{\lambda}\right|-\kappa\phi\right]}
  \nonumber \\
 &\qquad &  \qquad
    + \frac{1}{\sqrt{4\pi|\v|}} \int_{0}^{\infty}dk\, \PsiWk(k) \,
      e^{ik\left[\ln\left|\frac{\v}{\lambda}\right|+\kappa\phi\right]}
\label{eq:PsiexpWdW-c}\\
  & \equiv &  \PsiWR(\v,\phi) + \PsiWL(\v,\phi).  
\label{PsiexpWdW-d}
\end{eqnarray}
\label{eq:PsiexpWdW}%
\end{subequations}
The orthogonal sectors of ``right-moving'' (in a plot of $\phi$ \emph{vs.\ 
}$\v$) and ``left-moving'' states clearly correspond to the expanding and 
contracting branches of the classical solutions, Eq.\ (\ref{eq:classtraj}).  
\emph{A priori}, note that $\PsiWkR(k) = \PsiWk(k) \ (k<0)$ and 
$\PsiWkL(k) = \PsiWk(k) \ (k>0)$ need not be in any way related in the 
Wheeler-DeWitt theory.

We now show that generic states in the loop quantized theory decompose into
symmetric superpositions of expanding and collapsing (right- and left-moving)
Wheeler-DeWitt universes at large volume.  This follows simply
from Eq.\ (\ref{eq:eskWdW}), which may be written
\begin{equation}
e^{(s)}_k(\v) \cong 
  z \sqrt{\lambda}\left\{ \eW_{+|k|}(\v)e^{+i\alpha(|k|)} +  
      \eW_{-|k|}(\v)e^{-i\alpha(|k|)}  \right\}
      \qquad   |\v| \gg 2\lambda|k|,
\label{eq:eskWdWz}
\end{equation}
where $z=\sqrt{2}$ for the normalization of Eq.\ (\ref{eq:eWdW}) appropriate
to the range $-\infty<\v<\infty$ of Eq.\ (\ref{eq:ipWdW}),%
\footnote{Since states $\PsiW(\v,\phi)$ are symmetric in $\v$, one is free to 
restrict instead to positive $\v$ only, in which case the $\sqrt{4\pi}$ in 
Eq.\ (\ref{eq:eWdW}) should be $\sqrt{2\pi}$ and $z=1$.
} %
and the $\sqrt{\lambda}$ is present because the $e^{(s)}_k(\v)$ are
dimensionless, whereas the $\eW_k(\v)$ are not.  
The factor of $z\sqrt{\lambda}$ can be understood as arising from the
difference between normalization of the Wheeler-DeWitt eigenfunctions on the
continuous range $-\infty < \v < \infty$ \emph{vs.\ }the normalization of loop
quantum states on an infinite lattice with step-size $4\lambda$.
(Compare Appendix B of Ref.\ \cite{kp10a}.)

The relationship expressed in Eq.\ (\ref{eq:eskWdWz}) has long been known in
loop quantum cosmology on the basis of both analytic and numerical arguments.
See, for example, Eq.\ (5.3) of Ref.\ \cite{aps:improved}, as well as Eq.\
(3.1) of Ref.\ \cite{kp10a} -- which also contains a careful analysis of the
convergence properties of this limit -- among many others.  Here we have
confirmed that the asymptotic behavior of the exact solutions agrees precisely
with these earlier arguments.

We know that the limit Eq.\ (\ref{eq:eskWdWz}) is valid when $|\v|\gg 
2\lambda|k|$, and more generally, that $e^{(s)}_k(\v)$ has support only in 
the wedge $|k| \lesssim |\v|/2\lambda$.  Eq.\ (\ref{eq:eskWdWz}) therefore 
holds inside this wedge of support but clearly breaks down near its boundary 
$|k|=2|n|$.  From Eq.\ (\ref{eq:Psiexpn}), quite generally
\begin{subequations}
\begin{eqnarray}
\Psi(\v,\phi) & = & \int_{-\infty}^{\infty} dk\ \Psik(k) e^{(s)}_k(\v) e^{i\omega_k\phi}
\label{eq:Psiexpnklim-a}\\
 & \cong & 
   \int_{-|\v|/2\lambda}^{|\v|/2\lambda} dk\ \Psik(k) e^{(s)}_k(\v) 
   e^{i\omega_k\phi}.
\label{eq:Psiexpnklim-b}
\end{eqnarray}
\label{eq:Psiexpnklim}%
\end{subequations}
It is noted in Ref.\ \cite{kp10a} that one must take care to draw conclusions
concerning the asymptotic behavior of states in sLQC based on that of the
eigenfunctions because the convergence of the sLQC basis $e^{(s)}_k(\v)$ to
that of the Wheeler-DeWitt theory is not uniform in $k$.  Nonetheless, we
argue that for a wide class of quantum states, there will be a well-defined
region depending on the state in which this approximation will hold for that
state.

Specifically, replacement of $e^{(s)}_k(\v)$ in the expression Eq.\
(\ref{eq:Psiexpnklim}) with its asymptotic form Eq.\ (\ref{eq:eskWdWz}) will
be valid for values of the volume (significantly larger than that) for which
the Fourier transform $\Psik(k)$ does not have significant support outside the
wedge at that volume.  Quantum states are normalized, so we know that
$\Psik(k)$ is square-integrable.  Because functions of compact support are
dense in 
$L^2(\Re)$, there is a dense set of states for which, for every state
$\Psi(\v,\phi)$ in this set, there is \emph{some} value of $|k|$ whose value
will in general depend on the state -- call it
$k_{\Psi}$ -- outside of which $\Psik(k)$ has no 
support.  Therefore, for a dense set of states in the quantum theory the
replacement Eq.\ (\ref{eq:eskWdWz}) will be a good approximation for $|\v|\gg
2\lambda|k_{\Psi}|$.  It is worth emphasizing, therefore, that the domain of
applicability of the large-volume approximation is \emph{dependent upon the
quantum state} through the support of $\Psik(k)$.

For states satisfying this condition and within that domain of applicability,
we may write
\begin{equation}
\Psi(\v,\phi) \cong z\sqrt{\lambda} \int_{-|\v|/2\lambda}^{|\v|/2\lambda} dk\ \Psik(k)
   \left\{ \eW_{+|k|}(\v)e^{+i\alpha(|k|)} + \eW_{-|k|}(\v)e^{-i\alpha(|k|)}  \right\}
   e^{i\kappa|k|\phi}.
\label{eq:PsiWdW}
\end{equation}
It will be seen shortly that the first term corresponds to a contracting
universe, and the second, expanding.  To begin, we note from Eqs.\
(\ref{eq:omegak}), (\ref{eq:Psiexpn}), and (\ref{eq:esksk}) that $\Psik(k)$ is
even, $\Psik(-k)=\Psik(k)$.  Consider the first term alone.  As Eq.\
(\ref{eq:PsiWdW}) applies only for values of the volume for which $\Psik(k)$
has negligible support for $|k| >|\v|/2\lambda$, we may extend the range of
$k$-integration to $-\infty < k < \infty$.  By separating the integral
$\int_{-\infty}^{\infty}dk = \int_{-\infty}^{0}dk + \int_{0}^{\infty}dk$ and
making the change of variable $k'=-k$ in the first, one quickly finds
%
\begin{subequations}
\begin{eqnarray}
z\sqrt{\lambda} 
 \int_{-\infty}^{\infty} dk\ \Psik(k)\, \eW_{|k|} (\v)e^{i\alpha(|k|)} e^{i\k|k|\phi} & = & 
  2z\sqrt{\lambda} \int_{0}^{\infty} dk\ \Psik(k)e^{i\alpha(|k|)} \eW_{k}(\v) e^{i\k|k|\phi}  
\label{eq:PsiWdWL-a}\\
 & = & \PsiL(\v,\phi).
\label{eq:PsiWdWL-b}
\end{eqnarray}
\label{eq:PsiWdWL}%
\end{subequations}
As in the Wheeler-DeWitt case, Eq.\ (\ref{eq:PsiexpWdW}), $\PsiL(\v,\phi)$ 
clearly corresponds to a contracting quantum universe, with equivalent 
Wheeler-DeWitt Fourier transform
\begin{equation}
\PsiWkL(k) = \Psik(k) e^{i\alpha(|k|)}  \qquad  k>0.
\label{eq:PsikL}
\end{equation}
In an exactly similar way, the second term in Eq.\ (\ref{eq:PsiWdW}) is
\begin{equation}
\PsiR(\v,\phi) \equiv  2z\sqrt{\lambda}
  \int_{-\infty}^{0} dk\ \Psik(k)e^{-i\alpha(|k|)} \eW_{k}(\v) e^{i\k|k|\phi},
\label{eq:PsiWdWR}
\end{equation}
which describes an expanding universe with equivalent Wheeler-DeWitt Fourier
transform
\begin{equation}
\PsiWkR(k) = \Psik(k) e^{-i\alpha(|k|)}  \qquad  k<0.
\label{eq:PsikR}
\end{equation}
Therefore, for a dense set of quantum states and within the domain of
applicability of the large volume approximation $|\v|\gg2\lambda|k_{\Psi}|$
for the state $\Psi(\v,\phi)$, we may always write
\begin{equation}
\Psi(\v,\phi) \cong \PsiR(\v,\phi)  + \PsiL(\v,\phi)  
\qquad 
\mbox{\scriptsize $
\begin{pmatrix}
\text{large}\\
\text{volume}  
\end{pmatrix}.
$}
\label{eq:PsiLR}
\end{equation}
Unlike the Wheeler-DeWitt case, however, $\PsiL$ and $\PsiR$ are not
independent.  In fact, owing to the symmetry of $\Psik(k)$ in the loop quantum
case, the equivalent Wheeler-DeWitt Fourier transforms are essentially the
same, having equal modulus $|\PsikL(-k)|=|\PsikR(k)|$ and a fixed phase
relation given by $\exp(i\alpha(|k|))$ between them.  (Note that expressions
equivalent to Eqs.\ (\ref{eq:PsikL}) and (\ref{eq:PsikR}) may also be found in
Ref.\ \cite{kp10a}.   These relations are central to the ``scattering'' 
picture of loop quantum cosmology developed in that reference.)

Thus, we have shown from the exact solution that in the sense given by Eq.\
(\ref{eq:PsiLR}), at sufficiently large volume a dense set of states in flat
scalar loop quantum cosmology may be written as \emph{symmetric}
superpositions of expanding and contracting universes.  This is an essential
feature of loop quantum cosmology, deeply connected with the fact that these
cosmologies are non-singular \cite{mbmmo09a} -- all states ``bounce'' with a
finite maximum matter density.

It was observed long ago from numerical solutions that semiclassical states
``bounce'' symmetrically, including the dispersions of these states
\cite{aps,aps:improved}.  Analytic bounds on the dispersions of \emph{all}
states in this model have been proved in Refs.\ \cite{cor-singh08a,kp10a}, in
which further discussion of constraints on the sense in which such states are
symmetric can also be found.  (In this regard see also Refs.\
\cite{livmb12a,mbmmo09a}.)  Here we have demonstrated the symmetry of generic
states (not just quasiclassical ones) at large volume, in the sense of Eq.\
(\ref{eq:PsiLR}), directly from the exact solutions.

\section{Critical density and the ultraviolet cutoff}
\label{sec:rhocrit}

A significant part of the interest in loop quantum cosmology has arisen from
the fact that loop quantization seems to robustly and generically resolve
cosmological singularities; see Ref.\ \cite{ashsingh11,boj11a} for recent
overviews.  This was noticed first in numerical results for semiclassical
states \cite{aps,aps:improved}, subsequently observed in many other models
(see e.g.\ Refs.\ \cite{apsv07a,bp08a}), and finally proved analytically for
all quantum states in the model described in this paper in Ref.\
\cite{acs:slqc}.  In that paper it is shown that the expectation value (and
hence spectrum) of the matter density $\hat{\rho}|_{\phi}$ is bounded above by
\begin{subequations}
\begin{eqnarray}
\rho_{\text{crit}} & = & \frac{\sqrt{3}}{32\pi^2\gamma^3}\frac{1}{Gl_p^2}
\label{eq:rhocrit-a}\\
 & \approx & 0.41\cdot\rho_p,
\label{eq:rhocrit-b}
\end{eqnarray}
\label{eq:rhocrit}%
\end{subequations}
where $\rhop$ is the Planck density and the value of the Barbero-Immirzi
parameter $\gamma \approx 0.2375$ inferred from black hole thermodynamics has
been used \cite{ash-lew04a}.

We argue here that the existence of a universal upper bound to the density may
be traced to the ultraviolet cutoff for values of $|k| \gtrsim |\v/2\lambda|$
on the eigenfunctions $e^{(s)}_k(\v)$.

Classically, the matter density when the scalar field has value $\phi$ is
given by the ratio of the energy in the scalar field to the volume,
\begin{equation}
\rho|_{\phi} = \frac{p_{\phi}^2}{2V|_{\phi}^2}.
\label{eq:rhoclassdef}
\end{equation}
In Ref.\ \cite{acs:slqc} it is argued that a suitable definition for the 
corresponding quantum mechanical observable is
\begin{equation}
\hat{\rho}|_{\phi} = \frac{1}{2} \hat{A}|_{\phi}^2,
\label{eq:rhoquantdef}
\end{equation}
where
\begin{equation}
\hat{A}|_{\phi} \equiv 
   \frac{1}{\sqrt{\smash[b]{\hat{V}|_{\phi}}}}\,\hat{p}_{\phi}\,
       \frac{1}{\sqrt{\smash[b]{\hat{V}|_{\phi}}}}.
\label{eq:Adef}
\end{equation}
Even though the spectrum of this operator is not yet known, an upper bound on 
the spectrum of $\hat{A}|_{\phi}$ places a bound on the spectrum of 
$\hat{A}|_{\phi}^2$, thence on $\hat{\rho}|_{\phi}$.  Ref.\ \cite{acs:slqc} 
then observed that
\begin{subequations}
\begin{eqnarray}
\expct{\hat{A}|_{\phi}}_{\Psi} & = & 
  \frac{\melt{\Psi}{\hat{A}|_{\phi}}{\Psi}}{\bracket{\Psi}{\Psi}}
\label{eq:Aexpct-a}\\
 & = & \frac{\melt{\chi}{\hat{p}_{\phi}}{\chi}}{\melt{\chi}{\hat{V}|_{\phi}}{\chi}},
\label{eq:Aexpct-b}
\end{eqnarray}
\label{eq:Aexpct}%
\end{subequations}
where 
$\ket{\chi}$ is defined through
$\ket{\Psi}=\sqrt{\smash[b]{\hat{V}|_{\phi}}}\ket{\chi}$.  Thus, the
expectation value of $\hat{A}|_{\phi}$ (in the state $\ket{\Psi}$) may be
expressed as the ratio of expectation values of the momentum and volume
(in the state $\ket{\chi}$).  Even though the spectrum of $\hat{p}_{\phi}$ is
not bounded, they go on to show analytically that this ratio is nonetheless
bounded above by $\sqrt{3/4\pi\gamma^2G}/\lambda$ for all states in the domain
of the physical observables, leading directly to the bound given by Eq.\
(\ref{eq:rhocrit}) on the density.

Alternately, one might choose to define
\begin{equation}
\hat{\rho}|_{\phi} = 
  \frac{1}{2} \frac{1}{\hat{V}|_{\phi}}\hat{p}_{\phi}^2\frac{1}{\hat{V}|_{\phi}}.
\label{eq:rhoquantdefalt}
\end{equation}
A similar argument then shows 
\begin{equation}
\expct{\hat{\rho}|_{\phi}}_{\Psi} = \frac{1}{2}
  \frac{\melt{\omega}{\hat{p}_{\phi}^2}{\omega}}{\melt{\omega}{\hat{V}|_{\phi}^2}{\omega}},
\label{eq:rhoexpctalt}
\end{equation}
where now $\ket{\Psi}= \hat{V}|_{\phi}\ket{\omega}$.  
For convenience, we will adopt this latter definition of the density in the
sequel.

\subsection{Heuristic argument}
\label{sec:rhocritheur}

We offer here a new perspective on the existence of a universal upper bound to
the density by arguing that it can be seen as a consequence of the linear
scaling of the ultraviolet cutoff in the $e^{(s)}_k(\v)$ with volume, $|k|
\lesssim |\v/2\lambda|$.  
We offer both a new proof of the existence of a critical density in this model in
the volume representation, %
as well as an heuristic argument that has a clear and intuitive interpretation,
making it a simple matter to calculate the value of the critical density
simply from the slope of the scaling of the ultraviolet cutoff.

In fact, heuristically speaking, using Eq.\ (\ref{eq:rhoexpctalt}) we see that
the UV cutoff $|k| \lesssim |\v/2\lambda|$ implies that
\begin{subequations}
\begin{eqnarray}
\expct{\hat{\rho}|_{\phi}}_{\Psi} & = & \frac{1}{2}
  \frac{\expct{\hat{p}_{\phi}^2}_{\omega}}{\expct{\hat{V}|_{\phi}^2}_{\omega}}
\label{eq:rhoexpctbndmoral-a}\\ 
 & \sim & \frac{1}{2} \frac{(\hbar\k|k|)^2}{\hat{V}|_{\phi}^2}
\label{eq:rhoexpctbndmoral-b}\\
 & \lesssim & \frac{1}{2}
   \left(\frac{\hbar\k}{2\lambda}\right)^2 \left(\frac{|\v|}{2\pi\gamma\lp^2|\v|}\right)^2,
\label{eq:rhoexpctbndmoral-c}%
\end{eqnarray}
\label{eq:rhoexpctbndmoral}%
\end{subequations}
identical to the rigorous bound on the density -- Eq.\ (\ref{eq:rhocrit}) --
found in Ref.\ \cite{acs:slqc}.
The linear scaling in the UV cutoff on the eigenfunctions
thus, in this heuristic way, leads directly to the existence of the universal
critical density.  In particular, the slope of the scaling gives the value of
the critical density correctly.

This ``moral'' argument is of course not rigorous since $\hat{p}_{\phi}$ and
$\hat{\v}|_{\phi}$ do not commute.  There is, however, an interesting
\emph{reason} the ``moral'' argument works: as discussed in Sec.\
\ref{sec:repop}, the scalar momentum $\hat{p}_{\phi}$ and volume
$|\hat{\v}|_{\phi}$ operators \emph{approximately} commute, again as a
consequence of the ultraviolet cutoff.  Thus, also as a consequence of the
ultraviolet cutoff, the operator $\hat{p}_{\phi}$ is, though its spectrum is
not bounded, in effect bounded on subspaces of fixed volume.
This leads immediately to the upper bound on the density, as in the ``moral'' 
argument above.

The intended meaning of these statements is the following.  Because of the
ultraviolet cutoff,
\begin{subequations}
\begin{eqnarray}
|\hat{p}_{\phi} e^{(s)}_{k}(\v)| & = & \hbar\k|k|\cdot |e^{(s)}_{k}(\v)|
\label{eq:pphiesbnd-a}\\
 & \lesssim & \hbar\k \left|\frac{\v}{2\lambda}\right|\cdot |e^{(s)}_{k}(\v)|.
\label{eq:pphiesbnd-b}%
\end{eqnarray}
\label{eq:pphiesbnd}%
\end{subequations}
Consider the subspace spanned by volume eigenstates with volume less than or 
equal to some $\Vmax$.  While this subspace is not strictly invariant under 
the action of the operator $\hat{p}_{\phi}$, it is \emph{approximately} so 
because the off-diagonal matrix elements $\melt{\v}{\hat{p}_{\phi}}{\v'}$ are 
strongly suppressed.  The norm of states restricted to this subspace is
$\lVert\chi\rVert_{\Vmax}^2=\sum_{|\v|\leq\Vmax}|\chi(\v)|^2$.  Then we have
\begin{subequations}
\begin{eqnarray}
\lVert\hat{p}_{\phi} e^{(s)}_{k}(\v)\rVert_{\Vmax}^2 & = & 
   (\hbar\k|k|)^2 \sum_{|\v|\leq\Vmax} |e^{(s)}_{k}(\v)|^2
\label{eq:pphiesbndV-a}\\
 & \lesssim &  (\hbar\k)^2 
\left|\frac{\Vmax}{2\lambda}\right|^2 \sum_{|\v|\leq\Vmax} |e^{(s)}_{k}(\v)|^2
\label{eq:pphiesbndV-b}\\
 & = & (\hbar\k)^2 
   \left|\frac{\Vmax}{2\lambda}\right|^2 \lVert e^{(s)}_{k}(\v)\rVert_{\Vmax}^2
\label{eq:pphiesbndV-c}%
\end{eqnarray}
\label{eq:pphiesbndV}%
\end{subequations}
and we can see that $\hat{p}_{\phi}$ is in effect bounded on subspaces of
volume less than a given value.  
%
Given Eq.\ (\ref{eq:Aexpct}) or (\ref{eq:rhoexpctalt}), this helps explain
why the heuristic argument above gives the correct value for the critical
density.

\subsection{Proof in the volume representation}
\label{sec:rhocritvolrep}

To complete this heuristic argument we offer a new, alternative proof of the
existence of a critical density in this model in the volume representation,
using the definition Eq.\ (\ref{eq:rhoquantdefalt}) for the density.  (The
original proof of Ref.\ \cite{acs:slqc} is in a different representation of
the quantum states and operators and employs the definition Eq.\
(\ref{eq:rhoquantdef}) for the density, though their proof works for either
definition.)

The action of the gravitational constraint $\Theta$ in the volume
representation, Eq.\ (\ref{eq:theta}), may be written
\begin{equation}
(\Theta\omega)(\v,\phi) = \frac{1}{2}\left(\frac{\k}{2\lambda}\right)^2
   \v^2\left\{\omega(\v,\phi)-\overline{\omega}(\v,\phi)\right\},
\label{eq:thetasimple}
\end{equation}
where
\begin{equation}
\overline{\omega}(\v,\phi) = \frac{1}{2}\left[
   \sqrt{\left|1+\frac{4\lambda}{\v}\right|}\left|1+\frac{2\lambda}{\v}\right|
      \omega(\v+4\lambda,\phi)   +
   \sqrt{\left|1-\frac{4\lambda}{\v}\right|}\left|1-\frac{2\lambda}{\v}\right|
      \omega(\v-4\lambda,\phi)
   \right]
\label{eq:chibar}
\end{equation}
is approximately the average of the values of $\omega$ on either side of the
volume $\v$.  In this notation,
\begin{subequations}
\begin{eqnarray}
\expct{\hat{p}_{\phi}^2}_{\omega} & = & \hbar^2 \melt{\omega(\phi)}{\Theta}{\omega(\phi)}
\label{eq:pphiexpctchibar-a}\\
 & = & \frac{1}{2} \left(\frac{\hbar\k}{2\lambda}\right)^2
        \sum_{\v}\left\{\v^2\omega(\v,\phi)^*\omega(\v,\phi) -
           \v^2 \omega(\v,\phi)^*\overline{\omega}(\v,\phi)  \right\}
\label{eq:pphiexpctchibar-b}\\
 & = & \frac{1}{2} \left(\frac{\hbar\k}{2\lambda}\right)^2
        \left\{\expct{\hat{\v}|_{\phi}^2}_{\omega} -
           \sum_{\v}\v^2 \omega(\v,\phi)^*\overline{\omega}(\v,\phi)  \right\}.
\label{eq:pphiexpctchibar-c}%
\end{eqnarray}
\label{eq:pphiexpctchibar}%
\end{subequations}
We wish to show this quantity is bounded above by $(\hbar\k/2\lambda)^2
\expct{\hat{\v}|_{\phi}^2}_{\omega}$, in accord with the ``moral'' argument of
Eq.\ (\ref{eq:rhoexpctbndmoral}).  To proceed, define
\begin{subequations}
\begin{eqnarray}
\omega'(\v,\phi) & = & \omega(\v,\phi) e^{i\frac{\v}{4\lambda}\pi}
\label{eq:chiprime-a}\\
 & = & \omega(\v,\phi) e^{in\pi},
\label{eq:chiprime-b}%
\end{eqnarray}
\label{eq:chiprime}%
\end{subequations}
where $\v=4\lambda n$.  Clearly $\omega'^*\omega'=\omega^*\omega$, so $\omega$
and $\omega'$ have the same norm in the inner product of Eq.\ (\ref{eq:ip}).
However,
\begin{subequations}
\begin{eqnarray}
\overline{\omega'}(\v,\phi) & = & \frac{1}{2}\left[
   \sqrt{\left|1+\frac{4\lambda}{\v}\right|}\left|1+\frac{2\lambda}{\v}\right|
      \omega'(\v+4\lambda,\phi)   +
   \sqrt{\left|1-\frac{4\lambda}{\v}\right|}\left|1-\frac{2\lambda}{\v}\right|
      \omega'(\v-4\lambda,\phi)
   \right]
\label{eq:chiprimebar-a}\\
 & = & \frac{1}{2}\left[
   \sqrt{\left|1+\frac{4\lambda}{\v}\right|}\left|1+\frac{2\lambda}{\v}\right|
      \omega(\v+4\lambda,\phi)  e^{i(n+1)\pi} +
   \sqrt{\left|1-\frac{4\lambda}{\v}\right|}\left|1-\frac{2\lambda}{\v}\right|
      \omega(\v-4\lambda,\phi)  e^{i(n-1)\pi}
   \right]
\label{eq:chiprimebar-b}\\
 & = & -\overline{\omega}(\v,\phi) e^{in\pi}.
\label{eq:chiprimebar-c}%
\end{eqnarray}
\label{eq:chiprimebar}%
\end{subequations}
Thus,
\begin{subequations}
\begin{eqnarray}
(\Theta\omega')(\v,\phi) =
& = & \frac{1}{2}\left(\frac{\k}{2\lambda}\right)^2
   \v^2\left\{\omega'(\v,\phi)-\overline{\omega'}(\v,\phi)\right\}
\label{eq:thetachiprime-a}\\
 & = & \frac{1}{2}\left(\frac{\k}{2\lambda}\right)^2
   \v^2\left\{\omega(\v,\phi)+\overline{\omega}(\v,\phi)\right\}e^{in\pi}.
\label{eq:thetachiprime-b}%
\end{eqnarray}
\label{eq:thetachiprime}%
\end{subequations}
Since $\Theta$ is a positive operator,%
\footnote{Strictly speaking, since $\omega'$ is not generally a solution to the
constraint except in regions where $\omega=0$, we should check that $\Theta$ is
positive on all functions normalizeable in the inner product of Eq.\
(\ref{eq:ip}).
} %
we find
\begin{subequations}
\begin{eqnarray}
\melt{\omega'(\phi)}{\Theta}{\omega'(\phi)}
& = & 
   \frac{1}{2}\left(\frac{\k}{2\lambda}\right)^2
   \sum_{\v}\v^2\omega(\v,\phi)^*\left\{\omega(\v,\phi)+\overline{\omega}(\v,\phi)\right\}
\label{eq:thetaexpctchiprime-a}\\
 & = & \frac{1}{2} \left(\frac{\k}{2\lambda}\right)^2
        \left\{\expct{\hat{\v}|_{\phi}^2}_{\omega} +
           \sum_{\v}\v^2 \omega(\v,\phi)^*\overline{\omega}(\v,\phi)  \right\}
\label{eq:thetaexpctchiprime-b}\\
 & \geq & 0.
\label{eq:thetaexpctchiprime-c}%
\end{eqnarray}
\label{eq:thetaexpctchiprime}%
\end{subequations}
Eqs.\ (\ref{eq:pphiexpctchibar}) and (\ref{eq:thetaexpctchiprime}) show that
the absolute value of the sum of off-diagonal terms,
$\sum_{\v}\v^2\omega^*\overline{\omega}$, is bounded above by the sum of the
diagonal terms, $\expct{\hat{\v}|_{\phi}^2}_{\omega}$. 
Thus, from Eq.\ (\ref{eq:pphiexpctchibar}) we see that
\begin{equation}
\expct{\hat{p}_{\phi}^2}_{\omega}  \leq
  \left(\frac{\hbar\k}{2\lambda}\right)^2 \expct{\hat{\v}|_{\phi}^2}_{\omega},
\label{eq:phiexpctbnd}
\end{equation}
as desired.  With the definition Eq.\ (\ref{eq:rhoexpctalt}) for the density,
then, the heuristic ``moral'' argument showing the relation between the slope
of the scaling of the UV cutoff and the value of the critical density is
supported by a direct calculation.

A parallel demonstration in the volume representation using the definition
Eq.\ (\ref{eq:rhoquantdef}) of the density would similarly show that the sum
of the off-diagonal terms in $\expct{\hat{p}_{\phi}}_{\chi}$ is bounded above
by the sum of the diagonal terms.  Though this can be plausibly argued on the
basis of the observations in Sec.\ \ref{sec:repop} that the
off-diagonal elements of $\hat{p}_{\phi}$   
in the volume representation are bounded above by the diagonal elements, and
strongly suppressed for elements connecting more than one step off the
diagonal -- and is of course known to be true because of the proof of Ref.\
\cite{acs:slqc} -- a proof entirely in the volume representation at the same
level of rigor as that possible for $\hat{p}_{\phi}^2$ is more difficult
because the matrix elements of $\hat{p}_{\phi}$ are so much more complicated.
The ``moral'' argument applies in either case.

\section{Discussion}
\label{sec:discuss}

Working from recent exact results for the eigenfunctions of the dynamical
constraint operator in flat, scalar loop quantum cosmology, we have
demonstrated the presence of a sharp momentum space cutoff in the
eigenfunctions that sets in at wave numbers $|k| = |\v/2\lambda|$ that may be
understood as an ultraviolet cutoff due to the discreteness of spatial volume
in loop quantum gravity.  Earlier numerical observations showing the onset of
a rapid decay in the eigenfunctions at small volume at a volume proportional
to the eigenvalue $\omega_k$ are thus confirmed analytically in this model.
We have argued that the existence of a maximum (``critical'') value of the
matter density $\rho|_{\phi} = p_{\phi}^2/2V|_{\phi}^2$ that is universal in
the sense that it is independent of the state can be viewed as a consequence
of the ultraviolet cutoff since the minimum volume and maximum momentum scale
in the same way.  This bound holds for generic quantum states in the theory in
the domain of the Dirac observables, not only states which are semiclassical
at large volume.

We have offered both an heuristic ``moral'' argument based on the scaling of
the UV cutoff, and a new direct proof in the volume representation.  While the
``moral'' argument for the critical density is not rigorous, it is physically
and intuitively clear, and enables the value of the critical density to be
calculated straightforwardly as in Eq.\ (\ref{eq:rhoexpctbndmoral}) once the
slope of the scaling of the cutoff is known.   Consistency with the bounds on 
the matrix elements of the physical operators set by the UV cutoff shows the 
overall coherence of these different points of view.

It is our hope that this perspective on the origin of the critical density
will have some use in the study of more complex models.  In particular, while
the dynamical eigenfunctions have been calculated analytically in this simple
model, it is probably too much to hope that this will be accomplished in most
other, more complicated, models.  Rigorous proofs of the existence and value
of a universal critical density may therefore be difficult to achieve in many
models beyond sLQC. Nevertheless, in all models it should be possible to study
solutions to the gravitational constraint numerically.  With the recognition
from Ref.\ \cite{acs:slqc} that the density is bounded by the ratio of the
expectation value of the momentum to the volume, we have argued here that the
existence of a universal critical density may be viewed as due to the linear
scaling of the ultraviolet momentum space cutoff in the eigenfunctions
$e^{(s)}_k(\v)$ with volume.  Therefore, in models in which analytical
solutions are not available, numerical evidence for the existence of an
ultraviolet cutoff in the eigenfunctions may nevertheless be employed to argue
robustly for the existence of an upper bound to the matter density for generic
quantum states in those models, and indeed, its precise value may be inferred
from the slope of the cutoff scaling.


The asymptotics enabling the demonstration of the ultraviolet cutoff in the
eigenfunctions also enabled us to extract analytically the large volume limit
of these eigenfunctions based on an analysis of the model's exact solutions.
The result, consistent with considerable prior work in the field based on
physical, analytical and numerical arguments, is that the eigenfunctions
approach a particular linear combination of the eigenfunctions for the
Wheeler-DeWitt quantization of the same physical model, with a precise
determination of the phase, as well as some understanding of the domain of
applicability of the approximation.  In turn, this allowed us to show that
generic quantum states in the theory approach symmetric linear combinations of
``expanding'' and ``contracting'' Wheeler-DeWitt universes
at large volume, no matter how non-classical those states may be.

\appendix* 




\begin{acknowledgments}

D.C. would like to thank Parampreet Singh for the 
discussions which led to this work  %
and for critical comments on an earlier version of the manuscript%
, and Marcus Appleby for helpful conversations.  D.C.\ would also like to
thank the Perimeter Institute, where much of this work was completed, for its
hospitality.  Research at the Perimeter Institute is supported by the
Government of Canada through Industry Canada and by the Province of Ontario
through the Ministry of Research and Innovation.

\end{acknowledgments}


\ifthenelse{\arxiv=1}{%
\bibliography{eLQC}
}{%
\bibliography{global_macros,../Bibliographies/master}%
}%

\end{document}